\documentclass[aps,prd,twocolumn,groupedaddress,showpacs]{revtex4}
\usepackage{graphics}

\usepackage{amssymb}
\usepackage{amsmath}
\usepackage{amsfonts}

\begin{document}


\title{The Gravito-Maxwell Equations of General Relativity\\
in the local reference frame of a GR-noninertial observer}


\author{Christoph Schmid}           
\email{chschmid@itp.phys.ethz.ch}
\affiliation{ETH Zurich, Institute for Theoretical Physics, 
8093 Zurich, Switzerland}

\date{\today}

\begin{abstract}
 We show that the acceleration-difference of neighboring freefalling
 particles (= geodesic deviation) measured in the local reference frame
 of a GR-{\it noninertial} observer is {\it not} given by the Riemann  
 tensor. With the gravito-electric field $\vec{\cal{E}}_{\rm g}$ of GR
 defined as the acceleration of freefalling quasistatic particles relative
 to the observer, $\mbox{div}\,\vec{\cal{E}}_{\rm g}$ measured in the
 reference frame of a GR-noninertial observer is different from the
 curvature $R^0_{\,\,0}$. We derive our exact, explicit, and simple
 gravito-Gauss law for $\mbox{div}\,\vec{\cal{E}}_{\rm g}$ in our new
 reference frame of a GR-noninertial observer with his LONB (Local
 Ortho-Normal Basis $\bar{e}_{\hat{a}}$) and his LONB-connections
 $({\omega}_{\hat{b}\hat{a}})_{\hat{c}}$ in his time- and 3-directions:
 the sources of $\mbox{div}\,\vec{\cal{E}}_{\rm g}$ are contributed by
 all fields including the GR-gravitational fields $(\vec{\cal{E}}_{\rm g},
 \vec{\cal{B}}_{\rm g})$. In the reference frame of a GR-{\it inertial}
 observer our gravito-Gauss law coincides with with Einstein's $R^0_{\,\,0}$
 equation, which does not have gravitational fields as sources.
 We derive the gravito-Amp\`ere law for $\mbox{curl}\,\vec{\cal{B}}_{\rm g}$,
 the gravito-Faraday law for $\mbox{curl}\,\vec{\cal{E}}_{\rm g}$, and
 the law for $\mbox{div}\,\vec{\cal{B}}_{\rm g}$. The densities of energy,
 momentum, and momentum-flow of GR-gravitational fields
 $(\vec{\cal{E}}_{\rm g}, \vec{\cal{B}}_{\rm g})$ are {\it local} observables,
 but they depend on the observer with his local reference frame:
 if measured by a GR-inertial observer on his worldline in his frame of
 LONB connections, these quantities are zero. For a GR-noninertial observer
 the sources of gravitational energy, momentum, and momentum-flow densities
 have the opposite sign from the electromagnetic and matter sources. The
 sources in the gravito-Gauss law contributed by gravitational energy
 and momentum-flow densities have a {\it repulsive} effect on the
 gravitational acceleration-difference of particles. 
\end{abstract}


\maketitle


\section{Introduction, Method,  and Results
  \label{summary.x}}

The {\it gravito-Gauss law} of GR
gives $\mbox{div}\,\vec{\cal{E}}_{\rm g},$
where the gravito-electric field $\vec{\cal{E}}_{\rm g}$
is identical with the gravitational acceleration 
of quasistatic particles:
$\vec{\cal{E}}_{\rm g} \equiv
\vec{\mathfrak{g}}_{\rm quasistat}^{\,(\rm GR)}$.

But {\it acceleration} cannot be measured
without a {\it reference frame}:
  Newton's accelerations are defined and measured
  relative to a Newton-inertial frame.
  In classical mechanics accelerations are defined and measured
  also relative to reference frames which are
    accelerated and/or rotating relative to Newton-inertial.

    In GR,  gravitational acceleration
    ($\equiv$ acceleration of freefalling particles)
    is defined and measured {\it relative} to a {\it chosen observer},
    $\vec{\mathfrak{g}}_{\rm quasistatic}^{\,(\rm GR)}
    \equiv \vec{\cal{E}}_{\rm g}$.  
    A GR-{\it inertial} observer is freefalling and nonrotating
    relative to spin-axes of gyroscopes comoving on his worldline,
    and he measures $\vec{\cal{E}}_{\rm g} = 0.$
    Einstein 1911: "gravitational acceleration is relative,
    it depends on the observer"
    \cite{Einstein.1907,
          Einstein.happiest.thought,
          Einstein.1911,
          Einstein.1916}.~---
    The difference between
    a GR-inertial versus noninertial observer's reference frame
    is crucial in the gravito-Maxwell equations
    but irrelevant in Einstein's equations.

    The worldline of a free-falling test-particle
    defines a geodesic.  
    The acceleration-difference
    of neighboring free-falling test-particles
    is equal to the {\it geodesic deviation}.
    The measured geodesic deviation
    depends on the local reference frame
    of the chosen observer which includes
    his neighboring auxiliary observers.

We now prove that the {\it geodesic deviation}
measured in the local reference frame of
      a GR-{\it noninertial observer}
      is {\it not given} by the {\it Riemann tensor}.
      This is evident for the case of a vanishing Riemann tensor:
      the acceleration-difference of
      neighboring freefalling quasistatic particles
measured in a {\it rotating reference frame} 
(relative to gyrospin axes)
is non-zero because of centrifugal accelerations.
Hence the {\it geodesic deviation}
measured in a {\it rotating reference frame}  
is {\it non-zero} even if the {\it Riemann tensor} is {\it zero}.
GR-texts~\cite{Wald, Weinberg, Poisson.Will, Hartle}
are wrong in concluding the opposite.

In our {\it method}
the {\it reference frame} of the {\it observer} is crucial.
{\it Cartan's} method  works with a field of
{\it Local Ortho-Normal Bases}, LONBs~$\bar{e}_{\hat{a}} (P)$,
which we identify with a field of observers.
The LONB-components ${\cal{E}}_{\rm g}^{\,\hat{i}} \equiv
{\mathfrak{g}}_{\rm quasistat}^{\hat{i}}$  are 
measured by these observers on their worldlines. 
 Hats on indices denote LONB-components.
      The gravito-magnetic field is defined and measured
      by the gravitational spin-precession of
      the observer's gyroscopes   
      ${\cal{B}}_{\rm g}^{\,\hat{i}}\equiv
      -{\Omega}_{\rm gyros}^{\hat{i}}$
      relative to the observer's LONB. 

\begin{itemize}
\item Our {\it new method} uses our 
  {\it reference frame} of {\it LONBs}
  for a {\it noninertial observer}
      (accelerated and spinning relative to GR-inertial)
      with his Ricci
      {\it LONB-connections}~$(\omega^{\hat{b}}_{\,\,\hat{a}})_{\hat{c}}$
      in his {\it time- and 3-space-directions},
      Eqs.~(\ref{Ricci.connections.spat.local.frame}).
      The LONB-connections $(\omega^{\hat{b}}_{\,\,\hat{a}})_{\hat{i}}$
      in his 3-space directions  connect 
      to his auxiliary observers,
      which are co-accelerated, co-spinning, and co-orbiting.
    \end{itemize}
    {\it No coordinate-frame} exists
        adapted to a GR-rotating observer
        with his co-spinning and co-orbiting auxiliary observers
        as shown in Sect.~\ref{spinning.ff.prim.obs}.   

        Measured in the reference frame of an observer
        with his LONB-connections 
      $(\omega^{\hat{b}}_{\,\,\hat{a}})_{\hat{i}}$
      to his auxiliary observers,
      Eqs.~(\ref{Ricci.connections.spat.local.frame}):
      the acceleration difference of neighboring
      quasistatic freefalling particles,
      the {\it geodesic deviation},
divided by separations and {\it spherically averaged}
gives the 3-divergence of gravitational accelerations,
\begin{eqnarray}
  \mbox{div}\,\vec{\mathfrak{g}}_{\rm quasistat}^{\,(\rm GR)}\,
  &\equiv&\,\mbox{div}\,\vec{\cal{E}}_{\rm g}\,\,
         =\,\,\partial_{\hat{i}}{\cal{E}}_{\rm g}^{\hat{i}}.
           \nonumber
\end{eqnarray}

\begin{itemize}
\item
  Measured in the {\it reference frame} of LONBs
  for a GR-{\it noninertial} observer
with his LONB connections
$(\omega^{\hat{b}}_{\,\,\hat{a}})_{\hat{c}}$,
the spherical average of {\it geodesic deviations},
$\mbox{div}\,\vec{\cal{E}}_{\rm g}$,
is {\it different} from  the
Ricci curvature $R^{\,\hat{0}}_{\,\,\,\hat{0}}$.
Our new {\it exact result}, explicitely simple: 
\begin{eqnarray} R^{\,\hat{0}}_{\,\,\,\hat{0}}\,
  &=&\,\mbox{div}\,\vec{\cal{E}}_{\rm g}
      - (\vec{\cal{E}}_{\rm g}^{\,2} + 2 \vec{\cal{B}}_{\rm g}^{\,2}).
\label{R00.divE}
\end{eqnarray}
\end{itemize}
This equation gives the 
correction for incorrect conclusions on geodesic deviations in  
GR-books~\cite{Wald, Weinberg, Poisson.Will, Hartle}.

On the right-hand side of Eq.~(\ref{R00.divE})
every term depends on whether the observer
(with his frame of LONBs)
is GR-inertial or not.
In contrast, the Riemann tensor and its two parts, the Ricci and Weyl tensors,
do not depend on whether the chosen observer is GR-inertial or non-inertial.

  Subtracting Eq.~(\ref{R00.divE})
  from Einstein's $R^{\,\hat{0}}_{\,\,\,\hat{0}}$ equation,
  Eq.~(\ref{Einstein.R00.eq}), gives our new {\it exact result}:
  \begin{itemize}
    \item
    our {\it gravito-Gauss law}
    in an observer's  
    {\it frame} with LONB-connections,
    Eqs.~(\ref{Ricci.connections.spat.local.frame}),
\begin{eqnarray}
   \mbox{div}\,\vec{\cal{E}}_{\rm g}
   &=& - 4\pi G (\rho_{\varepsilon} + 3 \tilde{p})_{\rm matter}
       - G (\vec{E}^2 + \vec{B}^2) \nonumber
  \\
  &&\quad\quad
       +\,  (\vec{\cal{E}}_{\rm g}^{\,2} + 2 \vec{\cal{B}}_{\rm g}^{\,2}).
  \label{grav.Gauss}
\end{eqnarray}
Sources of $\mbox{div}\,\vec{\cal{E}}_{\rm g}$ are contributed
by {\it all fields}: matter, electromagnetic~$(\vec{E}, \vec{B})$,
{\it gravitational}~$(\vec{\cal{E}}_{\rm g}, \vec{\cal{B}}_{\rm g})$ with
$\rho_{\varepsilon} =$  energy density, 
$3\tilde{p} =$ trace of momentum flow density. 

The sources of $\mbox{div}\,\vec{\cal{E}}_{\rm g}$ contributed
by the gravitational $(\rho_{\varepsilon}+3\tilde{p})_{\rm grav}$
have  the {\it opposite sign}   from
the matter and electromagnetic sources
and have a {\it repulsive} effect on the 
gravitational acceleration-difference of test particles
measured by~$\mbox{div}\,\vec{\cal{E}}_{\rm g}$.
\end{itemize}

In the frame of a GR-{\it noninertial observer} 
our gravito-Gauss law, Eq.~(\ref{grav.Gauss}),
differs from Einstein’s $R^{\hat{0}}_{\,\,\hat{0}}$ equation,
\begin{eqnarray}
  R^{\hat{0}}_{\,\,\hat{0}}\,
  &=&\,-4\pi G (\rho_{\varepsilon}+3 \tilde{p})_{\rm matter}
      - G (\vec{E}^2 + \vec{B}^2).
      \label{Einstein.R00.eq}
\end{eqnarray}
In Einstein's equations the sources do  {\it not}  include  the
densities of {\it gravitational} energy, momentum, and momentum-flow:
the {\it symmetry} of the three sources (matter, electromagnetic, 
gravitational) is {\it missing}.

In our gravito-Gauss law of GR,
$\mbox{div}\,\vec{\cal{E}}_{\rm g}$ and
$ (\vec{\cal{E}}_{\rm g}^{\,2} + 2 \vec{\cal{B}}_{\rm g}^{\,2})$
{\it depend} on the  {\it acceleration} and {\it rotation}
(relative to GR-inertial) of the chosen local {\it observer}
with his reference frame of LONBs,
Eqs.~(\ref{Ricci.connections.spat.local.frame}).~---
In crucial  contrast, all terms in Einstein’s equations
are {\it independent} of
the acceleration and rotation
of the observer with his reference frame.~---
In the frame and on the worldline of a GR-{\it inertial observer},
$R^{\hat{0}}_{\,\,\hat{0}} = \mbox{div}\,\vec{\cal{E}}_{\rm g}$
and $\vec{\cal{E}}_{\rm g}^{\,2} = 0, \,\vec{\cal{B}}_{\rm g}^{\,2} = 0$,
and our gravito-Gauss law becomes {\it identical} with
Einstein's $R^0_{\,\,\,0}$
equation from Eqs.~(\ref{R00.divE})-(\ref{Einstein.R00.eq}).

But local frames of GR-{\it inertial} observers
(freefalling and nonrotating)
are {\it never} used
in the solar system nor for our inhomogeneous universe:

\begin{itemize}
\item {\it new result}: in our strongly inhomogeneous universe
  the gravito-Gauss law in the local reference frame of LONBs
  of a {\it non-inertial} observer
predicts {\it repulsive gravity} from 
$(\rho_{\varepsilon}+3\tilde{p})_{\rm grav}
= -(\vec{\cal{E}}_{\rm g}^{\,2} +2\vec{\cal{B}}_{\rm g}^{\,2})/(4\pi G)$.
This contributes to the measured
{\it accelerated expansion} of our universe today.
It is important to determine the magnitude of this effect.
\end{itemize}

We derive our three remaining exact, explicit, and simple
gravito-Maxwell equations of GR
in the local reference frame of LONBs for a GR-noninertial observer:
the gravito-Amp\`ere and Faraday laws
and the law for $\mbox{div}\,\vec{\cal{B}}_{\rm g}$,
\begin{eqnarray}
  \mbox{curl}\,\vec{\cal{B}}_{\rm g}
  &=&-8\pi G\,(\vec{J}_{\varepsilon})_{\rm matter}
      - 2G\,(\vec{E}\times\vec{B})_{\rm EM}
      \nonumber
      \\
  &&\quad
     + 2\,(\vec{\cal{E}}_{\rm g}\times\vec{\cal{B}}_{\rm g}),
 \nonumber
  \\
  \mbox{curl} \, \vec{\cal{E}}_{\rm g}
  &=&-\,2\,\partial_{\hat{t}}\,\vec{\cal{B}}_{\rm g},
      \nonumber
  \\
  \mbox{div}\,\vec{\cal{B}}_{\rm g}
  &=&-\,\vec{\cal{E}}_{\rm g}\cdot\vec{\cal{B}}_{\rm g}.
\end{eqnarray}
The gravito-Amp\`ere law of GR
has {\it no term} $\partial_{\hat{t}} \vec{\cal{E}}_{\rm g}$.

\begin{itemize}
\item
  The sources of $\mbox{div}\,\vec{\cal{E}}_{\rm g}$
and $\mbox{curl}\,\vec{\cal{B}}_{\rm g}$
include  terms which must be identified with the densities of 
gravitational energy~$\rho_{\varepsilon}^{(\rm grav)}$,
gravitational energy current~$\vec{J}_{\varepsilon}^{\,(\rm grav)}$,
and the trace of the gravitational
momentum-flow~$3\tilde{p}_{\rm grav}$
measured on the worldline and in the  local frame of LONBs 
for the {\it observer},
\begin{eqnarray} (\rho_{\varepsilon}+3\tilde{p})_{\rm grav} \,
&=&\,-\,(\vec{\cal{E}}_{\rm g}^{\,2}+2\vec{\cal{B}}_{\rm g}^{\,2})/(4\pi G),
    \label{grav.energy}
  \\
  \vec{J}^{\,(\varepsilon)}_{\rm grav}\,
  &=&\,-\,(\vec{\cal{E}}_{\rm g}\times \vec{\cal{B}}_{\rm g})/(4\pi G).
   \nonumber 
\end{eqnarray}
\item
  Our {\it new result}:
  our gravitational~$(\rho_{\varepsilon}+3\tilde{p})_{\rm grav}$
and $\vec{J}_{\varepsilon}^{\,(\rm grav)}$ are {\it local observables},
but they depend on the local {\it observer}
with his {\it reference frame} of LONBs.
These observables are {\it zero} if measured
by a GR-{\it inertial observer} on his worldline
and with his reference frame.
\end{itemize}

In dramatic contrast Landau and
Lifshitz~\cite{Landau.Lifshitz.class.fields.1951}
gave the gravitational energy density in arbitrary coordinates: 
in a (3+1)-split it has thousands of terms. 

Using Eq.~(\ref{grav.energy}) we can rewrite the gravito-Gauss law,
\begin{eqnarray}
   \mbox{div}\,\vec{\cal{E}}_{\rm g}
   &=& - 4\pi G (\rho_{\varepsilon} + 3 \tilde{p})_{\rm matter+EM+gravit}.
  \label{grav.Gauss.3}
\end{eqnarray}

In Sect.~\ref{R00.not.rel.accel} we prove
from Einstein's concepts 1911 (without using Einstein's equations of 1915)
that the source
of the gravito-Gauss law
contributed by the gravitational fields is 
$(\vec{\cal{E}}_{\rm g}^2 + 2\vec{\cal{B}}_{\rm g}^2)$,
hence repulsive
as in Eq.~(\ref{grav.energy}).

For Einstein's equations it is irrelevant,
whether the observer with his frame is GR-inertial or non-inertial.
But for the gravito-Maxwell equations of GR
with their gravitational energy, momentum, and momentum-flow
it is crucial, whether the observer with his frame
is GR-inertial or not.

The accelerated expansion of our {\it inhomogeneous} universe
is quantified theoretically by the divergence of
gravito-electric accelerations $\mbox{div}\,\vec{\cal{E}}_{\rm g}$
given by the gravito-Gauss law and integrated over our Hubble volume.
The sources of  $\mbox{div}\,\vec{\cal{E}}_{\rm g}$ include
the repulsive gravitational $\vec{\cal{E}}_{\rm g}^2$.

Very many papers on gravito-electromagnetism
have published equations for {\it linear} perturbations,
which is no substitute for our exact equations.

For exact treatments,
a very different approach from  ours
has been taken in most (maybe all) publications
on gravito-electromagnetism:
they work with {\it coordinate bases},
e.g. Ref.~\cite{JantzenCariniBini}.
But (a)~coordinate bases cannot be adapted
to a rotating observer
with his adapted auxiliary observers (co-orbiting and co-spinning)
as shown in our Sect.~\ref{spinning.ff.prim.obs}.
This fact and (b)~working with Riemann metric
functions~$g_{\mu\nu} (x^{\lambda})$
makes the (fully explicit)
gravito-Gauss equation highly nonlinear.

The simplicity of our exact and fully explicit
gravito-Gauss law, Eq.~(\ref{grav.Gauss}),
is due to our {\it local reference frame} of {\it LONBs}
for a GR-{\it noninertial observer}
with his Ricci {\it LONB connections}
$(\omega_{\hat{b}\hat{a}})_{\hat{c}}$
in his time-direction and his 3-directions, 
Eqs.~(\ref{Ricci.connections.spat.local.frame}).
This gives our exact  
and simple gravito-Gauss law:
the derivative term is {\it linear},
and the source contributed by gravitational fields 
is $(\vec{\cal{E}}_{\rm g}^{\,2} + 2\vec{\cal{B}}_{\rm g}^{\,2})$,
which has a {\it repulsive} effect.

\subsection{Concepts and methods for deriving
  \protect\\
  our gravito-Maxwell equations of GR
  \protect\\
in observer's reference frame of LONBs}

Our paper is based on:
(1) the crucial role of the observer,
GR-inertial versus non-inertial,
with his local reference frame of neighboring LONBs,
(2)~Cartan's method with his field of LONBs,
which we identify with a field of observers.~---
Our concepts and methods are different
from those presented in GR-texts and most research papers. 
Our entirely new methods are:

(A)~our Ricci LONB-connections $(\omega^{\hat{b}}_{\,\,\hat{a}})_{\hat{c}}$
in the time-direction and the 3-directions of a non-inertial observer,
Eqs.~(\ref{Ricci.connections.spat.local.frame}), which are directly
given by $(\vec{\cal{E}}_{\rm g}, \vec{\cal{B}}_{\rm g})$,

(B)~our Golden Rule, Eq.~(\ref{golden.rule.curvature}),
for the Riemann tensor in LONB-components
in terms of the Ricci LONB
connections~$(\omega^{\hat{b}}_{\,\,\hat{a}})_{\hat{c}}$
in the local frame of a non-inertial primary observer.

In Sect.~\ref{grav.el.field.x} 
we show how the gravito-electric
field~$\vec{\cal{E}}_{\rm g}^{\,(\rm GR)}$
is operationally defined and measured by 
a chosen {\it observer} on his worldline:
$\vec{\cal{E}}_{\rm g}^{\,(\rm GR)}$~is identical with the
gravitational acceleration~$\vec{\mathfrak{g}}_{\rm GR}$ of
quasistatic  particles {\it relative}
to the {\it observer}  measured on his worldline,
\begin{eqnarray}
  \vec{\cal{E}}_{\rm g}^{\,(\rm GR)}\,
  &\equiv&\,
    \vec{\mathfrak{g}}_{\,\rm quasistatic}^{\,(\rm rel.to\,obs)}.
      \nonumber
      \end{eqnarray} 
For a GR-inertial  observer 
$\vec{\cal{E}}_{\rm g}^{\,(\rm GR)} = 0$
measured by him on his worldline.  
``One can no more speak of absolute acceleration'' (Einstein).

In Sect.~\ref{grav.magn.field.x}
we give our new operational definition of the
gravito-magnetic field~$\vec{\cal{B}}_{\rm g} \,$
measured by an observer on his worldline:
the gravitational {\it precession}~$\vec{\Omega}$
of {\it gyroscopes} relative to the observer-LONB gives
$\vec{\cal{B}}_{\rm g}^{\,(\rm GR)}$,
\begin{eqnarray}
  \vec{\cal{B}}_{\,\rm g}^{\,(\rm GR)}\,
  &\equiv&\,-\,\vec{\Omega}_{\rm gyro-precession}^{(\rm rel.to\,obs.LONB)}.
           \nonumber
\end{eqnarray}
A GR-inertial observer measures 
$\vec{\cal{B}}_{\rm g}^{\,(\rm GR)} = 0$ on his worldline.~---  
GR-{\it inertial motion} of an observer or particle
is freefalling and non-rotating
relative to spin-axes of gyroscopes on its worldline.

Sect.~\ref{Ricci.conn.x} gives our new result that 
the Ricci LONB-connection~$(\omega_{\hat{a}\hat{b}})_{\hat{0}}$
of an observer's LONB   
along his worldline  
is given directly by the gravitational
fields~${\cal{F}}_{\hat{a}\hat{b}}^{(\rm g)}$
measured by the observer on his worldline,
\begin{eqnarray}
  (\omega_{\hat{i}\hat{0}})_{\hat{0}}\,
  &=&\,- {\cal{E}}_{\hat{i}}^{(\rm g)},
      \nonumber
  \\
  (\omega_{\hat{i}\hat{j}})_{\hat{0}}\,
  &=&\, - {\cal{B}}_{\hat{i}\hat{j}}^{(\rm g)}\,\,\equiv\,\,
     -\varepsilon_{\hat{i}\hat{j}\hat{k}} {\cal{B}}^{(\rm g)}_{\hat{k}},
     \nonumber
  \\
(\omega_{\hat{a}\hat{b}})_{\hat{0}}\,
  &=&\,- {\cal{F}}_{\hat{a}\hat{b}}^{(\rm g)}.
      \nonumber
\end{eqnarray}
Inside the bracket of~$(\omega_{\hat{a}\hat{b}})_{\hat{c}}$
are Lorentz transformation indices,
outside the bracket is the displacement index.   

In Sect.~\ref{Erlangen} we treat 
``Geometry without metric'': affine geometry, 
affine connections, and parallel transport. 
The Ricci LONB-connection
{\it along} the {\it worldline} of an {\it inertial observer}  
with his LONB carried along is,
\begin{eqnarray} \mbox{inertial observer:}\,\,\,\,\,
  (\omega^{\hat{b}}_{\,\,\hat{a}})_{\hat{0}} \,
  &=&\,0.   \nonumber
\end{eqnarray}
The LONB-connections in {\it radial} directions
from an {\it inertial} primary observer
to his adapted {\it auxiliary observers}
at infinitesimal $\delta{r}$
are given by auxiliary observers at rest and
nonrotated relative to the primary observer,
\begin{eqnarray}
  \mbox{frame of inertial observer:}\,\,\,\,\,
  (\omega^{\hat{b}}_{\,\,\hat{a}})_{\hat{k}}^{(r=0)}\,
  &=&\, 0.   \nonumber
\end{eqnarray}
Our frame of Ricci LONB-connections
of an inertial observer 
is our {\it new  tool}, fundamentally {\it different} from
the local inertial frame in GR-texts,
which is independent of the choice of an observer.~---
Our new frame of LONBs is crucial for our remarkable result
of Eq.~(\ref{Newton.Gauss.Einstein}).

Sect.~\ref{natural.aux.obs.x}
gives the construction of our most crucial new tool,
our local {\it reference frame} of LONBs for
a {\it non-inertial observer}
with his neighboring LONBs.
The LONB-connections for displacements
from a non-inertial observer
in his {\it time-direction} and {\it radial directions} are:
\begin{eqnarray}
     (\omega_{\hat{i} \hat{0}})_{\hat{0}}
     \,\,=\,\,-\,{\cal{E}}_{\hat{i}},\,\, 
     \,\,&&\,\,
     (\omega_{\hat{i} \hat{j}})_{\hat{0}}
     \,\,=\,\,-\,{\cal{B}}_{\hat{i}\hat{j}},
          \nonumber        
\\
  (\omega_{\hat{i}\hat{0}})_{\hat{j}}
  \,\,=\,\,-\,{\cal{B}}_{\hat{i} \hat{j}}, 
  \,\,&&\,\,
       (\omega_{\hat{i} \hat{j}})_{\hat{k}}
       \,\,=\,\,0.
       \nonumber
\end{eqnarray}
These LONB-connections, Eqs.~(\ref{Ricci.connections.spat.local.frame}),
are the key to our exact, explicit, and simple gravito-Maxwell equations.  

Sect.~\ref{eqs.of.motion.noninert.obs.xxx}
gives our GR-equations of motion for particle momenta and spins in 
our 
{\it reference frame} of LONBs
for a {\it non-inertial} observer:
exact, explicit, 
simple, 
\begin{eqnarray}
  (\frac{d}{d\hat{t}})_{\rm prim.obs}\,\,p^{\hat{i}}\,
     &=&\,   
       \varepsilon\, (\, \vec{\cal{E}}_{\rm g}
+ \,  2\vec{v} \times \vec{\cal{B}}_{\rm g} \, )^{\hat{i}} 
\,\,\,\,\,\mbox{gravit. forces} 
\nonumber
\\ 
&+&\, 
q\,(\,\vec{E}\,\, 
+\,\vec{v}\times\vec{B}\,)^{\hat{i}}, 
\,\,\,\,\mbox{el.mag. forces}.   
\nonumber
\end{eqnarray}

Sect.~\ref{sec.curvature.Cartan}
gives a brief review of Cartan's method for computing Riemann curvature
from the deficit LONB Lorentz transformation
after a round trip along an infinitesimal 
coordinate plaquette $[{\mu}, {\nu}].$
This gives Cartan's curvature 2-form equation,
\begin{eqnarray}
  (\tilde{\cal{R}}^{\hat{a}}_{\,\,\,\hat{b}})_{\mu\nu}\,
  &=&\,(d\,\tilde{\omega}^{\hat{a}}_{\,\,\,\hat{b}}\,\,
        + \,\,\tilde{\omega}^{\hat{a}}_{\,\,\,\hat{s}} \wedge
        \tilde{\omega}^{\hat{s}}_{\,\,\,\hat{b}})_{\mu\nu}.
      \nonumber
\end{eqnarray}
Cartan's LONB-method is unavoidable for computing Riemann curvature
from an observer's frame of LONB-connections.

In Sect.~\ref{ff.accel.diff.inert.obs}
we consider the geodesic deviation,
the freefall acceleration-difference,
spherically averaged,
in  our {\it frame} of LONBs
for an {\it inertial observer}, Eq.~(\ref{conn.inert.obs}):
Cartan's wedge terms vanish,
 $(\tilde{\omega}^{\,\hat{b}}_{\,\,\,\hat{c}}\wedge
   \tilde{\omega}^{\,\hat{c}}_{\,\,\,\hat{a}})_{\mu\nu} = 0,$    
  and the {\it exact} curvature is {\it linear} in
  $(\omega^{\hat{b}}_{\,\,\,\hat{a}})_{\hat{c}}$.  
  In our frame of a GR-{\it inertial observer},
for nonrelativistic physics
and neglecting electromagnetic fields,
       the {\it partial differential equation}
         for $R^{\,\hat{0}}_{\,\,\,\hat{0}}$
         is {\it identical} with (and follows from)
         the {\it Gauss law} for {\it Newton-gravity},
\begin{eqnarray}
R^{\,\hat{0}}_{\,\,\,\hat{0}}\,
  &=&\,
      \mbox{div}\,\vec{\mathfrak{g}}_{\rm GR}^{(\rm quasistat)}\, 
         =\,\mbox{div}\,\vec{g}_{\rm Newton}\,
=\, -4\pi G\rho_{\rm m}.
\nonumber
\end{eqnarray}
In Sect.~\ref{Newton.Gauss.to Einstein} we prove
that Einstein's equations follow from the Gauss-Newton
law $\mbox{div}\,\vec{g}_{\rm Newton} = -4\pi G\rho_{\rm mass}$
and the GR-concept of timelike geodesics (= freefall)
plus Special Relativity and the contracted Bianchi identity.~---
The gravito-Gauss law of GR 
in the frame of a GR-{\it inertial observer} is,
\begin{eqnarray}
\mbox{inertial obs:}\,\,\,\,\,\,\mbox{div}\,{\cal{E}}_{\rm g}\,
  &=&\, - 4\pi G (\rho_{\varepsilon} + 3\tilde{p})_{\rm matter + EM},
      \nonumber
\end{eqnarray}
where $(3\tilde{p}) \equiv$ trace of 3-momentum-flow tensor.

In Sect.~\ref{sect.golden.rule}   
we derive our Golden Rule,
Eq.~(\ref{golden.rule.curvature}),
for the Riemann tensor in LONB-components  
in terms of the Ricci LONB-connections
$(\omega^{\hat{a}}_{\, \, \, \hat{b}})_{\hat{c}}$
in the local frame of LONBs of a non-inertial primary observer.

In Sects.~\ref{gravito.Gauss.x}-\ref{div.grav.magn}
we use this Golden Rule
to derive our exact gravito-Maxwell equations 
in our local {\it frame} of LONB-connections
$(\omega^{\hat{a}}_{\, \, \, \hat{b}})_{\hat{c}}$
for a {\it non-inertial observer}.~---
In our gravito-Gauss law the {\it repulsive} gravitational
source $(\rho_{\varepsilon} + 3 \tilde{p})_{\rm grav}
= - (\vec{\cal{E}}_{\rm g}^{\,2} + 2\vec{\cal{B}}_{\rm g}^{\,2})/(4\pi G)$
contributes to the {\it accelerated expansion}
of our 
{\it inhomogeneous universe}.

Sect.~\ref{R00.not.rel.accel} 
is based on Einstein's concepts of 1911
without Einstein's equations of 1915:
we give an elementary re-derivation of our exact and new
{\it repulsive} source term
$(\vec{\cal{E}}_{\rm g}^{\,2} + 2\vec{\cal{B}}_{\rm g}^{\,2})$
in our gravito-Gauss law 
for~$\mbox{div}\,\vec{\cal{E}}_{\rm g}$.  

\section{Gravito-electric and
  \protect\\
  gravito-magnetic fields
of GR}
\label{gravit.fields}

\boldmath
\subsection{GR-gravito-electric field identical with
  \protect\\
  gravitional acceleration of quasistatic particles
  \protect\\
  relative to observer
\label{grav.el.field.x}}
\unboldmath

We show how in GR the gravito-electric field $\vec{\cal{E}}_{\rm g}$
is operationally defined and measured:
$\vec{\cal{E}}_{\rm g}$ is identical to the gravitational
acceleration $\vec{\mathfrak{g}}$ of quasistatic particles.
The gravitational acceleration
is measured {\it relative} to the chosen
{\it observer} on his worldline:
measured by a GR-inertial observer on his worldline
$\vec{\cal{E}}_{\rm g} = 0$.

We start from the definition of
the classical Maxwell-electric field~$\vec{E}$ 
for negligible gravity:

(1)~via the electric force $\vec{F}_{\rm el}$
on a quasistatic test-particle
with infinitesimal charge,
$\, \lim_{q \rightarrow 0} \, (\vec{F}_{\rm el}/q)~\equiv~\vec{E},\,$

(2)~with Newton's second law,
$\,\vec{F} = m_{\,\rm inertial} \,\vec{a},\,$
and the acceleration $\,\vec{a}$ measured relative to 
an inertial frame of Newton resp. 
Special Relativity.

Hence the electric field $\vec{E}$ is defined and measured by:
\begin{eqnarray}
       \vec{E}\,\,
     &\equiv\,\,(m_{\rm inertial}/q)\,\,\,
     \vec{a}_{\,\rm particle\,released\,from\,rest,only\,ED}
       ^{\,(\rm rel.to\,inertial\,Newton\,resp.SR)}.
       \label{def.Maxwell.el.field}
\end{eqnarray}
The test-particle must be quasistatic, otherwise
the acceleration by the magnetic Lorentz force
would also contribute, $\,\vec{a}_{\,\rm Lorentz}
= (q/m)\, [\,\vec{v} \times \vec{B}\,].$~---
The electric field~$\vec{E}$,
measured with quasistatic particles,
is used in the equations of Special Relativity
for relativistic particles.

Starting from the definition of the Maxwell-electric field
in Eq.~(\ref{def.Maxwell.el.field}),
three steps are needed for the definition of
the gravito-electric field of GR:

1) Replace electric charge $q$ in the electric force
by gravitational mass $m_{\,\rm gravit}$  
       in the gravitational force on quasistatic particles.
        Write $m_{\,\rm inertial}$ in Newton's 2nd  law. 

        2) Use that gravitational acceleration
        is independent of the substance, 
        tested by E\"ot-Wash to $\pm 10^{-13}$~\cite{EotWash}: 
        the ratio of $ (m_{\rm gravit} / m_{\rm inert}) $
        is a universal constant,
        and units are chosen to make it equal to~1.~---
Going from electrodynamics to gravitodynamics of quasistatic particles,
$(q/m)\,\Rightarrow\,(m_{\rm gravit}/m_{\rm inertial})\,=\,1.$

3) Recognize that the definition of acceleration by
        $\vec{a} \equiv (d\vec{v}/dt)$ alone is {\it empty}:
        accelerations must be measured
        {\it relative} to  a {\it reference frame}.

          Newton's gravitational accelerations
          are defined and measured
          in Newton-inertial frames, they are {\it absolute}.
          Einstein's gravitational
          accelerations are defined and measured
          {\it relative} to the frame of the chosen local observer.

These three steps give the general operational definition
of the gravito-electric field in GR,
\begin{eqnarray}
        \vec{\cal{E}}_{\rm g}^{\,(\rm GR)}  
  &\equiv&\,\vec{\mathfrak{g}}_{\rm quasistat}^{\,(\rm GR)}
           \,\equiv\,
           \vec{a}_{\,\rm freefall,quasistat}^{\,(\rm rel.to\,local\,obs)}.
\label{def.grav.E.xxx}        
\end{eqnarray}
A {\it free-falling observer} measures on his worldline:
\begin{eqnarray}
        \vec{\cal{E}}_{\rm g}^{\,(\rm GR)}  
   &\equiv&\,\vec{\mathfrak{g}}_{\rm quasistat}^{\,(\rm GR)}
            \,=\,0.
            \nonumber
 \end{eqnarray}
 Einstein's gravitational acceleration $\vec{\mathfrak{g}}$
 is {\it relative},
 it depends on the chosen local {\it observer's acceleration
 relative to freefall}.
Einstein's ``happiest thought of my life'' in 1907
~\cite{Einstein.1907, Einstein.happiest.thought,
       Einstein.1911, Einstein.1916}:   
  ``The gravitational field has only a relative existence: ... 
  for an observer falling freely from the roof of a house,
  there exists, 
  at least in his immediate surroundings,
  no gravitational field''.

    GR-textbooks (and many papers) do not give a general definition of
    gravitational acceleration and gravito-electromagnetic fields:
    their gravitational acceleration and gravito-electromagnetic fields
    are in the frame of observers at rest relative to the Newton-inertial
    frame for the solar system. The gravitational accelerations
    in GR-textbooks are {\it absolute} in contradiction to Einstein's
    ``happiest thought of my life'': for free-falling observers, who measure
    zero gravitational acceleration on their worldlines, GR-textbooks
    are wrong.~---
    Inhomogeneous cosmology has no unique field of preferred local
    frames/observers, therefore gravitational accelerations and
    gravito-electromagnetic fields are necessarily relative, they depend on
    the chosen space-time slicing with observers at rest on that slicing.

        The components of the gravito-electric
        field~${\cal{E}}_{\rm g}^{\,\hat{i}} \equiv 
        {\mathfrak{g}}^{\hat{i}}_{\rm quasistat}$  
are defined and measured in the chosen observer's
Local Ortho-Normal Basis, LONB~$\bar{e}_{\hat{a}}$
in the tangent spaces $TM_P$ 
for points $P$ on his worldline.
  {\it Hats} on indices denote a LONB,
  Misner, Thorne, and Wheeler, MTW~\cite{MTW}.
  Our sign convention for the Minkowski metric
  is from MTW~\cite{MTW},
  $\eta_{\hat{a}\hat{b}} = \mbox{diag}\,(-1,+1,+1,+1).$~---
In $\mathfrak{a}^{\hat{i}} = (dv^{\hat{i}}/d\hat{t}\,)$  
the time-interval is measured on the observer's chronometer. 

Einstein's equations alone do {\it not predict}
  gravitational accelerations and the gravito-electromagnetic
  fields: the choice of a field
  of {\it observers} (GR-inertial vs. non-inertial)
  with their {\it LONBs} is needed to predict
  gravitational accelerations and gravito-electromagnetic fields.

To first order in  $\delta \hat{t}$,
a particle released by the observer into freefall
will acquire a first-order velocity change~$\delta v^{\hat{i}}$,
but the particle will still be
on the observer's worldline,
since its displacement will be of second order in~$\delta \hat{t}.$~---
{\it No role} for {\it Riemann metrics} 
in tangent spaces with Minkowski metric
on the worldline of the observer. 

Test-particles must be quasistatic,
otherwise the gravito-Lorentz
acceleration~$\vec{\mathfrak{g}}_{\rm GR} \propto
(\vec{v}\times\vec{\cal{B}}_{\rm g})$  contributes.

  In GR  the gravitational
  acceleration ${\mathfrak{g}}^{\hat{i}}_{\rm quasistat}
  \equiv {\cal{E}}_{\rm g}^{\hat{i}}$
  cannot be a 3-vector, because it vanishes
  in the reference
  frame of a freefalling observer.
  Hence we use the mathfrak-notation~${\mathfrak{g}}^{\hat{i}}$
  and the cal-notation~${\cal{E}}_{\rm g}^{\hat{i}}$.
  We show in Sect.~\ref{sect.E.B.Ricci.x} that
  ${\mathfrak{g}}^{\hat{i}}_{\rm quasistat} 
  \equiv {\cal{E}}_{\rm g}^{\hat{i}}  
  = - (\omega^{\,\hat{i}}_{\,\,\,\hat{0}})_{\hat{0}},$ 
the observer's Ricci LONB-connection
for a displacement along his worldline.

The gravito-electric field 
$\vec{\cal{E}}_{\rm g}$
measured by the acceleration of
freefalling quasistatic test particles  
relative to the observer
is valid in relativistic equations of motion  
and  field equations in the local frame of this observer.  

    \subsubsection{Einstein-equivalence of  classical
      fictitious accelerations
      \protect\\
      with GR-gravitational accelerations
      without matter sources
         \label{sect.fict.acc}}
 
Measuring accelerations needs a {\it reference frame}.
In contexts where Newton gravity and classical mechanics hold, 
if the acceleration of a freefalling particle
is measured relative to a Newton-{\it non-inertial} frame,
there are fictitious forces and {\it fictitious accelerations},
\begin{eqnarray}
   &&\vec{\mathfrak{g}}_{\,\rm GR, quasistatic}^{\,(\rm rel.to\,GR-obs)}
      \,\,=
     \nonumber
     \\
   &&\vec{g}_{\,\rm Newton}^{\,(\rm rel.to\,Newton-inert)}
     + (\vec{a}_{\,\rm class.fict})_{\rm Newton-inert}
      ^{(\rm rel.to\,GR-obs)}.
      \label{g.Einstein.minus.g.Newton}
\end{eqnarray}
Einstein's gravitational
acceleration~$\vec{\mathfrak{g}}_{\,\rm GR}$
       is the sum of~$\vec{g}_{\,\rm Newton}$,
       which is generated by {\it mass sources},
       plus the {\it classical fictious acceleration}
       of the Newton-inertial frame relative
       to the chosen GR-observer, which is
       {\it not generated} by {\it mass sources},
       Eq.~(\ref{g.Einstein.minus.g.Newton}).

       Since GR has no Newton-inertial frames, 
       one must use Einstein's {\it equivalence} between
       classical {\it fictitious accelerations}
       of freefalling particles
       relative to a Newton-{\it non-inertial} frame 
       with  {\it contributions}
       to~$\vec{\mathfrak{g}}_{\,\rm GR}$
       {\it without matter sources} :
     ``Two relatively accelerated systems $K$ and $K'$
  have an equal title as systems of reference.
  We are able to     {\it produce} a {\it gravitational field}
  merely by {\it changing the system of reference}''.\,
  Einstein, 1916~\cite{Einstein.1916}.

In Einstein's definition of $\vec{\mathfrak{g}}_{\rm GR}$, 
the observer is arbitrary.
But natural {\it fields} of ``Fiducial Observers''  
are chosen as discussed in the next subsection.

  \subsubsection{Field of Fiducial Observers:  FIDOs     
  \label{subsect.FIDOs}}

  {\it Cartan's} method uses
  a field of Local Ortho-Normal Bases, LONBs,
  which we identify with a field of 
  {\it observers} each with his  LONB
  in the tangent spaces $TM_P$ at points on his world-line. 

There is great freedom in choosing Cartan's LONBs. 
  But one chooses a field of
  {\it ``Fiducial Observers''}, FIDOs, 
  a concept introduced by Thorne et al~\cite{Thorne.black.holes}.
  Examples:

  (1)~For the solar system with planets as test-particles,
      a field of adapted FIDOs is singled out:
      every FIDO remains at fixed measured distance from the Sun and
       at fixed $(\theta, \phi)$ and with LONBs non-spinning
      relative to the perihelia of outer planets.
    These GR-Fiducial Observers for the solar system are Einstein-noninertial,
    they are identical with Newton-inertial observers of the solar system,
    and they are at rest and nonspinning in Schwarzschild coordinates. 

(2)~For our {\it inhomogeneous universe}
a field of cosmic FIDOs is singled out:
  relative to the CMB-sky with the CMB-dipole removed,
  (a)~our FIDOs remain fixed in angular position
  and orientation of their
  $(\bar{e}_{\hat{\theta}}, \bar{e}_{\hat{\phi}})$,
(b)~FIDOs have an isotropic expansion rate 
given by the angular average
of the measured expansion rate of galaxies
in a bin of luminosity distance.   
These cosmic FIDOs have an {\it unperturbed Hubble expansion}.~---
These cosmic FIDOs are  
needed to operationally define and measure the gravito-electric
field~${\cal{E}}_{\rm g}^{\hat{i}}$ of GR in our
inhomogeneous  universe.

\boldmath
\subsection{GR-gravito-magnetic field
  $\vec{\cal{B}}_{\rm g}$
  \protect\\
  measured by gravitational precession
$\vec{\Omega}_{\rm gyro}
^{\,(\rm relat.obs)}$
\label{grav.magn.field.x}}
\unboldmath

The gravito-magnetic field~$\vec{\cal{B}}_{\rm g}$
of GR is operationally defined and measured by the
gravitational precession of gyroscopes
{\it relative} to the {\it observer's} LONB
with gyros comoving on 
observer's worldline.
If measured by a GR-inertial observer on his worldline
$\vec{\cal{B}}_{\rm g} = 0$.~---
Our operational definition of~$\vec{\cal{B}}_{\rm g}$
is new.  

Measuring~$\vec{\cal{B}}_{\rm g}$ is analogous to
measuring the Maxwell-magnetic
field~$\vec{B}$ by the precession of magnetic dipoles
relative to inertial frames of Newton or Special Relativity.

The spin  and the magnetic dipole moment~$\mu_{\rm mag}$ 
point in  different directions in general, e.g. for the Earth.
For GR instructive is 
a classical magnetic dipole with 
{\it identical constituent-distributions} of {\it charges}
and {\it masses},
\begin{eqnarray}
     \vec{\mu}_{\,\rm mag}\,
  &=&\,\frac{q}{2 m} \, \vec{S},
  \nonumber
\end{eqnarray}
and the angular velocity of spin-precession,
\begin{eqnarray}
  \vec{\Omega}_{\rm precession}\,
  &=&\,-\,\frac{q}{2m}\,\vec{B}.
\label{Omega.B}
\end{eqnarray}

The transition from the 
definition of~$\vec{B}$ in classical magnetism
to our operational definition
of the gravito-magnetic field $\vec{\cal{B}}_{\rm g}$
of General Relativity needs four steps:

(1)~We replace the  electromagnetic coupling
to the electric charge~$q$ 
by the gravitational  coupling  
to the nonrelativistic gravitational mass~$m_{\rm grav}$. 

(2)~We use the result
that the ratio of gravitational mass to inertial mass
is universal within experimental errors $\pm 10^{-13}$,
and  units are chosen such that this ratio is set to 1.
Hence, going from electrodynamics to gravitodynamics,
$(q/m) \Rightarrow (m_{\rm grav}/m_{\rm inertial}) = 1,$
\begin{eqnarray}
     \vec{\mu}_{\,\rm grav.magn}  \,\,
  &\equiv& \, \, \, \frac{1}{2} \, \vec{S}.
     \nonumber
\end{eqnarray}

     (3)~The gyro-precession must be
     measured relative to a reference frame.
     The only reference frame available in GR is
     the reference frame of the chosen local observer
     with his LONB in the tangent spaces on his world line.

     (4) Our {\it normalization}
       of~$\vec{\cal{B}}_{\rm g}$ is  
       fixed by  $({\cal{E}}_{\hat{i}}^{\,(\rm g)}, \, 
     {\cal{B}}_{\hat{i}}^{\,(\rm g)})$
     transforming under a change of the observer's acceleration
     and rotation as components of the gravitational
       field~${\cal{F}}_{\,\hat{a}\hat{b}}^{\,(\rm g)}$,
     Sect.~\ref{sect.E.B.Ricci.x}.~---
     Our normalization of $\vec{\cal{B}}_{\rm g}$ differs
     from two unjustified normalizations in the literature:
     (a)~requiring the GR-field equation
     for the gravito-magnetic field to have
     the same prefactor   
     for~$\vec{J_{\varepsilon}}$
     as in the classical Amp\`ere law for~$\vec{J}_{q}$,
     (b)~requiring the equation of motion
     for particle-motion to have
     the same prefactor  
     for~$\vec{\cal{B}}_{\rm g}^{\,(\rm GR)}$
     as in the classical
     Lorentz-acceleration~$\,\vec{a}_{\,\rm Lorentz}
     = q\,\vec{v}\times\vec{B}.$

     Our four steps give the {\it operational definition} 
     of the {\it gravito-magnetic field} of GR  
     by the gyro-precession angular velocity
     of two non-aligned gyros
     (comoving on observer's worldline)
      relative to the observer-LONB,
     \begin{eqnarray}
     ({\cal{B}}^{\,\hat{i}}_{\,\rm g})_{\,\rm GR}\,
     &\equiv&\,-\,
    ({\Omega}^{\hat{i}})_{\,\rm precess.comoving\,gyros}
                  ^{\,(\rm rel.to\,local\,obs-LONB)}.
      \label{def.B.g.x}
\end{eqnarray}
The Hodge-duals 
${\cal{B}}_{\hat{i}\hat{j}}^{(\rm g)}$
and $\Omega_{\hat{i}\hat{j}}$
are given via the totally antisymmetric
Levi-Civita 3-tensor~$\varepsilon_{\hat{i} \hat{j} \hat{k}}$ with
$\varepsilon_{\hat{1} \hat{2} \hat{3}} \,
  \equiv \, + 1$,
\begin{eqnarray} 
  {\cal{B}}_{\hat{i} \hat{j}}\,\,
  \equiv\,\,  
\varepsilon_{\hat{i} \hat{j} \hat{k}} \, {\cal{B}}_{\hat{k}} 
\,&\Rightarrow&\, 
  {\cal{B}}_{\hat{1} \hat{2}} \, \,
  \equiv \, \, {\cal{B}}_{\hat{3}},
\nonumber
\\
  (\frac{dS^{\hat{f}}}
  {d\hat{t}_{\rm obs}})^{(\rm rel.obs.LONB)}_{\rm gyro\,precession}
  &=&-
  (\Omega^{\hat{f}}_{\,\,\hat{i}})^{(\rm rel.obs.LONB)}_{\rm gyro\,precession}
      S^{\hat{i}}.
\label{dS.dt.Omega.indices}
\end{eqnarray}

Gravito-magnetism is {\it simpler} than classical magnetism,
where macroscopic bodies
have non-identical constituent-motions of charges versus masses. 
In gravito-magnetism the constituent-motions
of $m_{\rm   gravit}$ and
$m_{\rm inertial}$ are identical.

A single observer {\it cannot use}
the gravito-Lorentz acceleration
$\vec{\mathfrak{a}}_{\rm Lorentz}
\propto \vec{v} \times \vec{\cal{B}}_{\rm g}$  
to measure the gravito-magnetic field {\it on his worldline}.
Measuring the gravito-Lorentz acceleration
needs test-particles with finite velocity,
and acceleration measurements need a second velocity measurement
in a tangent space $TM (P_2)$ away from the primary observer's worldline.
This needs an appropriate choice of   
{\it auxiliary observers} with their LONBs, 
a task solved in our crucial 
Sect.~\ref{natural.aux.obs.x}
(where $\vec{\cal{B}}_{\rm g}$ is needed as an input).

The gravitomagnetic $\vec{\cal{B}}_{\rm g}$ of GR
has been measured by Foucault in Paris in 1853
with gyroscopes precessing relative to
his   
$\vec{e}_{\hat{i}} =$ (East, North, vertical).~---
The satellite-observatory Gravity Probe B
has measured $ \vec{\cal{B}}_{\rm g}$
by gyroscope precession
relative to  its local LONB 
$\, \bar{e}_{\hat{0}} = \bar{u}_{\rm GPB} \,$
and $\, \bar{e}_{\hat{i}} \, $
determined by the lines of sight to two quasars.

\subsection{Inertial motion in General Relativity}

One cannot remove the  pillar of Newton's inertial motion
from a theory of gravity without putting in its place
the new pillar of inertial motion in General Relativity.
  Einstein's revolutionary  concept
  of inertial motion of {\it point-particles} is free-falling motion.
\begin{itemize}
\item Operational definition of GR-inertial motion:

  1) {\it free-falling} motion,

  2) {\it non-rotating} motion  
relative to  spin-axes of
two comoving gyroscopes  with
spin-axes not parallel.
\end{itemize}
Many GR-texts do not give the second part of
the definition of GR-inertial motion: non-rotating. 

\boldmath
\section{Ricci LONB-connection  
  $(\omega^{\hat{b}}_{\, \, \, \hat{a}})_{\hat{0}}$
  \protect\\
  along worldline of observer
\label{Ricci.conn.x}}
\unboldmath

{\it Cartan} uses a field of Local Ortho-Normal Bases,
LONBs~$\bar{e}_{\hat{a}}$.
We identify Cartan's LONB-field with a field of {\it observers}.~---
We derive our new result that
$(\vec{\cal{E}}_{\rm g},\vec{\cal{B}}_{\rm g})$
measured by an observer on his worldline 
equals his Ricci
LONB-connection~$(\omega^{\hat{b}}_{\,\,\hat{a}})_{\hat{0}}$
along his worldline, Eq.~(\ref{EB.id.LONB.conn}).

Relative to GR-inertial motion,
the derivative of the observer's LONB along his worldline
is equal to the covariant derivative as discussed in
Sect.~\ref{aff.conn.worldline.inertial},
Eq.~(\ref{GR.abs.derivative}),
\begin{eqnarray}
  (\frac{d\bar{e}_{\hat{a}}}{d\hat{t}})_{\rm obs.LONB}
  ^{(\rm rel.to\,GR-inertial)}\,
  &=&\,(\frac{D\bar{e}_{\hat{a}}}{D\hat{t}})_{\rm obs.LONB}.
      \nonumber
\end{eqnarray}
The covariant derivative~$(D\bar{e}_{\hat{a}}/D\hat{t})$
of the observer's LONB along his worldline is given by the
Ricci LONB-connection of the observer's LONB
along his worldline, 
\begin{eqnarray}
  (\frac{D\bar{e}_{\hat{a}}}{d\hat{t}})_{\rm obs.LONB}\,
  &=&\,\bar{e}_{\hat{b}}^{(\rm obs)}\,
      (\omega^{\,\hat{b}}_{\,\,\,\hat{a}})_{\hat{0}}.
      \nonumber
\end{eqnarray}

Combining the last two equations to eliminate 
$(D\bar{e}_{\hat{a}}/D\hat{t})_{\rm obs.LONB}$ gives,
\begin{eqnarray}
   (\frac{d\bar{e}_{\hat{a}}}{d\hat{t}})_{\rm obs.LONB}
  ^{(\rm rel.to\,GR-inert)}\,
  &=&\,\bar{e}_{\hat{b}}^{(\rm obs)}\,
      (\omega^{\,\hat{b}}_{\,\,\,\hat{a}})_{\hat{0}}^{(\rm obs)}.
      \nonumber
\end{eqnarray}
The observer's acceleration and rotation
(Lorentz transformations)
relative to GR-inertial motion are equal to the Ricci
connection~$(\omega^{\hat{b}}_{\,\,\,\hat{a}})_{\hat{0}}$ 
of the observer's LONB along his worldline. 

\boldmath
      \subsection{Observer's LONB connection
        $(\omega^{\hat{i}}_{\,\,\hat{0}})_{\hat{0}} 
  = - {\cal{E}}_{\rm g}^{\hat{i}}$}
\unboldmath

In the observer's  LONB
$(du_{\rm obs}^{\hat{a}}/d\hat{t}_{\rm obs}) = 0.$~---
The covariant
derivative $(D\bar{u}_{\rm obs}/D\hat{t}_{\rm obs})^{\hat{i}}$ 
relative to the observer's LONB 
gives the  Ricci
LONB-connection~$(\omega^{\hat{i}}_{\,\,\hat{0}})_{\hat{0}}$
of the observer's LONB along his worldline,
\begin{eqnarray}
  (D\bar{u}_{\rm obs}/D\hat{t}_{\rm obs})^{\hat{i}}\,
  &=&\,(\omega^{\hat{i}}_{\,\,\hat{0}})_{\hat{0}}.
      \nonumber
\end{eqnarray}
%
$(D\bar{u}_{\rm obs}/D\hat{t}_{\rm obs})^{\hat{i}}$ 
  along the observer's worldline
is equal to the acceleration of the observer
relative to GR-{\it inertial} motion
as discussed in Sect.~\ref{aff.conn.worldline.inertial},
\begin{eqnarray}
     (\frac{du^{\hat{i}}_{\rm obs}}
      {d\hat{t}_{\rm obs}})^{(\rm rel.to\,GR.inertial)}\,  
      &=&\,  (\omega^{\hat{i}}_{\,\,\hat{0}})_{\hat{0}}.
     \nonumber
\end{eqnarray}

In Sect.~\ref{grav.el.field.x}
we have presented the opposite point of view:
the {\it acceleration} of a {\it freefalling} particle
{\it relative} to the {\it observer}
with his LONB  
gives the operational definition of the
gravito-electric field~$\vec{\cal{E}}_{\rm g}$,  
\begin{eqnarray}
  ({\cal{E}}_{\rm g}^{\hat{i}})^{(\rm rel.to\,obs)}\,
  &\equiv&\,({\mathfrak{a}}^{\hat{i}})_{\rm freefall\,from\,rest}
           ^{(\rm rel.to\,obs-LONB)}.
      \nonumber
\end{eqnarray}

Comparing the two points of view of the last two equations
(acceleration of what? relative to what?)
gives the relation between the gravito-electric
field~${\cal{E}}_{\rm g}^{\hat{i}}$
measured by the observer and the Ricci
LONB-connection~$(\omega^{\hat{i}}_{\,\,\hat{0}})_{\hat{0}}$
of the observer-LONB along his worldline, our new result,
\begin{eqnarray}
  (\omega^{\hat{i}}_{\,\,\hat{0}})_{\hat{0}}\,
  &=&\,-\,{\cal{E}}_{\rm g}^{\hat{i}}.
      \label{E.g.Ricci}
\end{eqnarray}

The LONB-connection  
$(\omega^{\hat{i}}_{\,\,\,\hat{0}})_{\hat{0}}
= - ({\mathfrak{g}}^{\hat{i}})^{(\rm GR)}_{\rm quasistat}
= - {\cal{E}}^{\hat{i}}_{\rm g}$      
is  {\it directly measurable} and depends on the {\it observer}.   

\boldmath
\subsection{Observer's LONB connection
  $(\omega_{\hat{j}\hat{i}})_{\hat{0}}
  = - {\cal{B}}_{\hat{j}\hat{i}}^{(\rm g)}$}
\unboldmath

For gyroscopes carried by the chosen observer, 
the gyro-spin precession relative to the observer-LONB
per measured observer-time is,
   \begin{eqnarray}
     (\frac{dS^{\hat{j}}}{d\hat{t}_{\rm obs}})_{\rm gyro\,precession}
     ^{(\rm rel.obs.LONB)}\,
     &=&\,-\,(\Omega^{\hat{j}}_{\,\,\,\hat{i}}
         \,S^{\hat{i}})_{\rm gyro\,precession}^{(\rm rel.obs.LONB)}.
         \nonumber
   \end{eqnarray}
   The covariant derivative of a gyro-spin is zero,
   \begin{eqnarray}
     ( \frac{ dS^{\hat{j}} }{ d\hat{t}_{\rm obs} } )
     _{\rm gyro\,precession}^{ (\rm rel.obs.LONB) }
     &=&-\,(\omega^{\hat{j}}_{\,\,\hat{i}})_{\hat{0}}\,\,\,
         (S^{\hat{i}})_{\rm gyro}^{(\rm rel.obs.LONB)}.
        \nonumber
   \end{eqnarray}
   The last two equations give the Ricci connection,
   \begin{eqnarray}
     (\omega_{\hat{j}\hat{i}})_{\hat{0}}\,
     &=&\,(\Omega_{\hat{j}\hat{i}})_{\rm gyro\,precession}
         ^{(\rm rel.obs.LONB)}.
         \nonumber
   \end{eqnarray}

Our  operational definition of the
gravito-magnetic field of GR has been given in  
Eq.~(\ref{def.B.g.x}),
\begin{eqnarray}
  {\cal{B}}^{(\rm g)}_{\hat{j}\hat{i}}\,
  &\equiv&\,-\,(\Omega_{\hat{j}\hat{i}})_{\rm precess.comov.gyros}
           ^{(\rm relat.to\,obs.LONB)}.
      \nonumber
\end{eqnarray}
Comparing the last two equations we obtain
the crucial relation between the gravito-magnetic
field~${\cal{B}}^{(\rm g)}_{\hat{j}\hat{i}}$
(measured by the observer on his worldline)  and
the Ricci
LONB-connection~$(\omega_{\hat{j}\hat{i}})_{\hat{0}}$
of the observer-LONB  
per displacement~$\bar{e}_{\hat{0}} = \bar{u}_{\rm obs}$,  
our new result,
\begin{eqnarray}
  (\omega_{\hat{j}\hat{i}})_{\hat{0}}\,
  &=&\,-\,{\cal{B}}_{\hat{j}\hat{i}}^{(\rm g)}.
      \label{B.omega}
\end{eqnarray}

\boldmath
\subsection{Gravitational fields 
  ${\cal{F}}_{\hat{b}\hat{a}}^{(\rm g)}$
  measured by observer:
  \protect\\
  Ricci connection of his
  LONB $(\omega_{\hat{b}\hat{a}})_{\hat{0}}$
\label{sect.E.B.Ricci.x}}
\unboldmath

The fields $({\cal{E}}_{\hat{j}}^{(\rm g)},
\,{\cal{B}}_{\hat{j} \hat{i}}^{(\rm g)})$
measured by the observer on his worldline
are combined in the gravitational
field~${\cal{F}}_{\hat{b}\hat{a}}^{(\rm g)}$,
which is equal to
minus the Ricci connection of the observer's LONB
for displacements along his worldline,
\begin{eqnarray}
  {\cal{F}}_{\hat{b}\hat{a}}^{(\rm g)}\,
  &=&\,-\, (\omega_{\hat{b}\hat{a}})_{\hat{0}},
      \nonumber
\\
  {\cal{E}}_{\hat{k}}^{\,(\rm g)}\,\,
  =\,\,{\cal{F}}_{\hat{k}\hat{0}}^{\,(\rm g)}\,
     &=&\,-\,(\omega_{\hat{k}\hat{0}})_{\hat{0}},
         \nonumber
  \\
  {\cal{B}}_{\hat{k}\hat{i}}^{\,(\rm g)}\,\,
  =\,\,{\cal{F}}_{\hat{k}\hat{i}}^{\,(\rm g)}\,
  &=&\,-\,(\omega_{\hat{k}\hat{i}})_{\hat{0}}.
      \label{EB.id.LONB.conn}
   \end{eqnarray}

The electromagnetic fields are related to the field
tensor~$F_{\hat{b}\hat{a}}$
in an identical way,
$ E_{\hat{k}} = F_{\hat{k}\hat{0}},\, 
B_{\hat{k}\hat{i}} = F_{\hat{k}\hat{i}}.$~--- 
Eqs.~(\ref{EB.id.LONB.conn}) are {\it new}
and fix our normalization of~$\vec{\cal{B}}_{\rm g}.$

A {\it  single observer} with his LONB in the Minkowski
tangent spaces~$TM_P$ for~$P$ on his worldline 
measures the acceleration of freefalling particles (released by him)
and the precession of gyroscopes (comoving on his worldline):
he measures his  gravitational
fields~$(\vec{\cal{E}}_{\rm g}, \vec{\cal{B}}_{\rm g})$
on his worldline and his Ricci LONB
connection~$(\omega_{\hat{b}\hat{a}})_{\hat{0}}$
along his worldline.
The gravitational fields~~$(\vec{\cal{E}}_{\rm g}, \vec{\cal{B}}_{\rm g})$
{\it depend} on the {\it observer}: they are zero
when measured by a GR-inertial observer on his worldline.---
In Minkowski tangent spaces on the worldline of an observer:
{\it no possible role} for {\it Riemann
metrics}
for  particular spacetime geometries.

\section{Geometry without
\protect\\
  Riemann metrics
\label{Erlangen}}

Affine connection with parallel transport in GR
needs affine geometry but not Riemann metrics.~---
Different space-time geometries have different Riemann metrics.

\subsection{Affine connection and parallel transport
\protect\\
 along worldlines 
 of GR-inertial observers
 \label{aff.conn.worldline.inertial}}

GR-{\it inertial motion} is identical with
GR-{\it parallel transport} and
GR-{\it affine connection}
in a timelike direction.

    The LONB-connection
    in a timelike direction  
    is given by GR-inertial motion of an observer: 

    1) {\it free-fall} gives a GR-straight worldline
        (geodesic)
        and GR-parallel transport 
        of~$\bar{e}_{\,\hat{0}}~\equiv~\bar{u}_{\rm obs}$,   

    2)  {\it non-rotating}, parallel transport of
         $\bar{e}_{\,\hat{i}}^{\,(\rm obs)},$
         is given experimentally by spin-axes
         of two comoving gyros
         with spin-axes not parallel, 
  \begin{eqnarray}
&&\mbox{on worldline of inertial observer:}
\nonumber
\\
    && \,\,\,
       (\omega^{\,\hat{b}}_{\,\,\,\hat{a}})_{\hat{0}}^{(\rm
       GR-inertial\,observer)}
      \,\, = \,\, 0.
       \label{conn.along.GR.inert.worldline}
  \end{eqnarray}

The  GR-covariant derivative
in a timelike direction~$\bar{e}_{\hat{0}}$
is the derivative relative to the LONB
of a  GR-{\it inertial observer}
with~$\bar{u}_{\rm obs}~=~\bar{e}_{\hat{0}}$,
    \begin{eqnarray}
       \nabla_{\,\hat{0}}^{\,(\rm GR)}\,
       &\equiv&\,
       \partial_{\,\hat{0}}^{\,(\rm rel.to\,GR-inertial\,obs.LONB)}.
       \label{GR.abs.derivative}
  \end{eqnarray}

{\it No Riemann metrics} are needed 
for the GR-affine connection  
of vectors and tensors in LONB components.    

\subsection{Local reference frame of LONBs
  \protect\\
  for GR-inertial observer}
\label{aux.obs.adapted.to.inertial.prim.obs.x}

Our new tool is the local     
{\it reference frame} of LONBs
for a primary {\it inertial observer}  
along his worldline, 
which includes a field of neighboring LONBs
(of adapted auxiliary observers)
at infinitesimal separations, Eq.~(\ref{conn.inert.obs}) below. 
This new frame makes our exact and explicit equations
remarkably simple.~---
Our new frame is entirely {\it different} from
the ``local inertial coordinate system'' in the literature,
which does not refer to a primary observer.

\subsubsection{Affine connection and  parallel transport for
  \protect\\
  radial displacements from primary inertial observer
\label{parallel.trsp.spacelike}}

For an {\it inertial} primary observer
with $\bar{u}_{\rm obs}^{(P)} = \bar{e}_{\hat{0}}^{(P)}$
and his 
LONB $\bar{e}_{\hat{i}}^{\,(P)}$ on his worldline,
we define  his {\it adapted auxiliary observers}
for 1st-order-infinitesimal
radial displacements~$\delta r$ 
by two requirements: 

(1) adapted auxiliary observers are at {\it relative rest}
with the primary inertial observer: 
$\bar{e}_{\hat{0}}^{\,(\rm aux.obs)}$ and
$\bar{e}_{\hat{0}}^{\,(\rm prim.obs)}$
are related by {\it parallel transport} under
infinitesimal radial displacements.  
The first-order radial distance
and 3-direction of an auxiliary observer
relative to $\bar{e}_{\hat{i}}^{\,(\rm prim.obs)}$ are
independent of time.
Hence the affine connection for {\it radial} displacements
from a primary {\it inertial} observer to his
adapted auxiliary observers at relative rest is, 
     \\
     \begin{eqnarray}
  (\omega^{\hat{i}}_{\,\,\,\hat{0}})_{\hat{k}}^{(r = 0)}\,&=&\,0.
     \nonumber
\end{eqnarray}

(2) The 3-LONBs $\,\bar{e}_{\hat{i}}$
of {\rm auxiliary observers}
and of the primary observer
must be  parallel under infinitesimal radial displacements,
i.e. relatively {\it non-rotated},
\begin{eqnarray}
 (\omega^{\hat{j}}_{\,\,\,\hat{i}})_{\hat{k}}^{(r = 0)}\,&=&\,0.
       \nonumber
       \end{eqnarray}

       The last two conditions for
       {\it radial} displacements from the worldline
       of an {\it inertial} primary observer
       plus the condition for a
       displacement {\it along} his {\it worldline} gives,
\begin{eqnarray}
\mbox{inertial primary observer:}
&&(\omega^{\hat{b}}_{\,\,\hat{a}})_{\hat{c}}^{(r = 0)}\,
=\,0.
\label{conn.inert.obs}                                             
\end{eqnarray}

      The four covariant derivatives~$\nabla_{\hat{a}}$
      in the LONB-directions~$\bar{e}_{\hat{a}}$
      of a primary GR-inertial observer
      are identical
  with the ordinary derivatives~$\partial_{\hat{a}}$
  in the the local frame of LONBs of the GR-inertial observer,
\begin{eqnarray}
     \nabla_{\hat{a}}^{(r = 0)}\,
 &=&\,\partial_{\hat{a}}^{\,(\rm rel.to\,local\,frame\,of\,GR-inert.obs)}.
     \label{covar.deriv.id.ord.deriv.Einstein.inert.frame}
      \end{eqnarray}      
{\it No role} for a {\it Riemann metric}.

\begin{itemize}
\item
  Our spatially local {\it frame} along the  
worldline of an {\it inertial} observer
with the Ricci LONB connections in {\it his time} and {\it his spatial
directions}~$(\omega^{\hat{b}}_{\,\,\hat{a}})_{\hat{c}}$
of Eq.~(\ref{conn.inert.obs})
is new and fundamentally {\it different} from  the
``local inertial frame'' of GR-texts,
which is independent of a primary observer.
\end{itemize}

We shall show that in  our reference
  frame of LONBs for a 
  GR-{\it inertial}
  observer, Eq.~(\ref{conn.inert.obs}),
  for nonrelativistic physics and without electromagnetism,
the {\it partial differential equation}
for $R^{\,\hat{0}}_{\,\,\,\hat{0}}$
is {\it identical} with the {\it Gauss law} for {\it Newton-gravity},
Sect.~\ref{Newton.Gauss.to Einstein},
\begin{eqnarray}
&&\,\,\,R^{\hat{0}}_{\,\,\hat{0}}
  =  \mbox{div}\,\vec{\mathfrak{g}}_{\rm GR}^{\,(\rm quasistat)} 
  = \mbox{div}\,\vec{g}_{\rm Newton}
= -4\pi G\rho_{\rm m}.
\nonumber
\end{eqnarray}

Note: along the worldline of a GR-{\it inertial primary}
  observer, his adapted {\it auxiliary} observers
  are {\it not} GR-{\it inertial} because
  of tidal forces and  relative torques.

  \section{Frame of non-inertial observer:
  \protect\\
  LONBs of neighboring observers
\label{natural.aux.obs.x}}

The crucial concept and new tool
for proving our exact, explicit, and simple
gravito-Maxwell equations of GR
in Sect.~\ref{gravito.Maxwell.x}
is the {\it local frame} of LONBs
for a {\it non-inertial primary observer}
with his neighboring LONBs,
which can be interpreted as LONBs of his
{\it adapted auxiliary observers},
Eqs.~(\ref{Ricci.connections.spat.local.frame}).
This frame of LONBs needs:

(A)~the Ricci LONB-connection in the
{\it time-direction} of the primary observer,
$(\omega^{\,\hat{b}}_{\,\,\, \hat{a}})_{\hat{0}},$
given in Eqs.~(\ref{E.g.Ricci},   \ref{B.omega}),
\begin{eqnarray}
(\omega_{\,\hat{i} \hat{0}})_{\hat{0}}  
 \,=\,-\,{\cal{E}}^{(\rm g)}_{\hat{i}},
&&\quad\,\, 
 (\omega_{\,\hat{i} \hat{j}})_{\hat{0}} 
\,=\,-\,{\cal{B}}^{(\rm g)}_{\hat{i}\hat{j}}, 
       \nonumber
\end{eqnarray}

(B) the LONB-connections for {\it radial displacements}
from the primary observer
to his {\it adapted auxiliary observers}  
at infinitesimal $\delta r$, 
  $(\omega^{\,\hat{b}}_{\,\,\,\hat{a}})_{\hat{i}}^{(r = 0)}$, derived now.

\subsubsection{Accelerated-nonspinning primary observer}

Our auxiliary observers,
at infinitesimal radial displacements  
from the primary observer and adapted
to the {\it accelerated-nonspinning} primary observer
satisfy two conditions:

(a) Adapted auxiliary observers are    
{\it at rest} relative to the primary observer
compared by parallel transport
in an infinitesimal radial displacement,
i.e. their $\bar{e}_{\hat{0}}$ are {\it radially parallel},
\begin{eqnarray}
     (\omega^{\,\hat{i}}_{\,\,\,\hat{0}})_{\hat{k}}^{(r = 0)}
     \,&=&\,0.
     \nonumber
\end{eqnarray}

(b) Adapted auxiliary observer have spatial LONBs
radially parallel  to
the primary observer's $\bar{e}_{\,\hat{i}}^{\,(r = 0)}$, 
\begin{eqnarray}
     (\omega^{\,\hat{j}}_{\,\,\,\hat{i}})_{\hat{k}}^{\,(r = 0)}
  \,&=&\,0.
        \nonumber
\end{eqnarray}
Hence the radial LONB-connections from    
the primary observer vanish, 
\begin{eqnarray}
&&\mbox{accelerated-nonspinning primary observer:} 
\nonumber
\\
&&\quad\quad\quad\quad\quad\,\, 
   (\omega^{\,\hat{b}}_{\,\,\, \hat{a}})_{\hat{k}}^{\,(r = 0)}
   \,\,=\,\,0.   
\label{Ricci.radial.accel.nonrot}  
\end{eqnarray}

\subsubsection{Spinning-freefalling primary observer}
\label{spinning.ff.prim.obs}

For a spinning and freefalling primary observer,
our adapted auxiliary observers at infinitesimal $\delta r$ are
in {\it circular co-orbiting} motion around  and {\it co-spinning}
with the primary spinning observer.
The {\it auxiliary worldlines} are {\it circular spirals}
winding around the primary worldline.

To first order in  $\delta r$,
the auxiliary observers have nonrelativistic velocities
relative to the primary observer,
\begin{eqnarray}
  (v_{\hat{i}})_{\rm aux.obs}^{(\rm rel.to\,prim.obs)}   
\,&=&\,      
      [\vec{\Omega}_{\,\rm prim.obs} \, \times \,
      \delta \vec{r}_{\rm aux.obs}]_{\hat{i}}
\nonumber  
\\
&=&\,-\,\Omega_{\,\hat{i}\hat{k}}^{\,(\rm prim.obs)}\,
\delta r_{\hat{k}}^{(\rm aux.obs)}.
\nonumber
\end{eqnarray}
The velocities of the auxiliary observers
relative to the primary observer give the Ricci LONB-connection
$(\omega_{\,\hat{i}\hat{0}})_{\hat{k}}^{(r = 0)}$
for radial displacements,
\begin{eqnarray}
  (\omega_{\,\hat{i}\hat{0}})_{\hat{k}}^{(r = 0)}\,
  &=&\,-\,(\Omega_{\,\hat{i}\hat{k}})_{\rm prim.obs}^{(\rm rel.to\,gyros)}
 \,\,\equiv\,\,-\,({\cal{B}}^{\,(\rm g)}_{\,\hat{i}\hat{k}})^{(r = 0)}.
      \nonumber
\end{eqnarray}
The LONB-vector
$\bar{e}_{\hat{0}}^{\, (\rm aux.obs)} =   \bar{u}_{\rm aux.obs}$
has an infinitesimal Lorentz-boost
relative to $\bar{e}_{\hat{0}}^{\, (\rm prim.obs)}$
in the direction of orbital motion.

Spatial LONB-vectors $\bar{e}_{\hat{i}}$
  of auxiliary observers
in direction of motion
have  corresponding infinitesimal Lorentz-boosts
in  the time-direction. 
They generate {\it spacelike spiral} worldlines
  winding around the primary observer's worldline. Therefore:
  \begin{itemize}
  \item
  {\it no fully adapted coordinates exist} for
  a rotating primary observer with his auxiliary observers
  (co-orbiting and co-spinning).
  \end{itemize}

There are no rotations of
3-LONBs  $\bar{e}_{\,\hat{i}}^{\,(\rm aux.obs)}$
for our adapted auxiliary observers
relative to the 3-LONBs of the primary observer,
\begin{eqnarray}
  (\omega_{\hat{i} \hat{j}})_{\hat{k}} \,
  &=& \, 0.
   \label{omega.i.j.k}
    \nonumber
\end{eqnarray}

The results above, 
first for an accelerated-nonrotating primary observer,
afterwards for a nonaccelerated-rotating primary observer, 
are {\it additive}.

Our crucial result:  
in the spatially local {\it reference frame} of LONBs
for a {\it noninertial} primary observer
(with his adapted auxiliary observers),
all 24~Ricci  LONB-connections   
for displacements along his worldline
and in his radial directions
are  either given by
$\vec{\cal{E}}_{\rm g}$ and $\vec{\cal{B}}_{\rm g}$
or zero:
\begin{itemize}
\item {\bf  frame of LONBs for noninertial observer:}
\boldmath
\begin{eqnarray}
  &&\mbox{displacement along worldline of primary observer,}
  \nonumber
  \\
  &&\quad\quad\quad\quad
     \quad
     (\omega_{\hat{i} \hat{0}})_{\hat{0}}^{(r = 0)}
     \,\,=\,\,-\,{\cal{E}}_{\hat{i}}^{\,(\rm g)}, 
     \nonumber
  \\
  &&\quad\quad\quad\quad
     \quad
     (\omega_{\hat{i} \hat{j}})_{\hat{0}}^{(r = 0)}
     \,\,=\,\,-\,{\cal{B}}_{\hat{i}\hat{j}}^{\,(\rm g)},
\nonumber
  \\
  &&\quad\mbox{radial displacement from primary worldline,}
     \nonumber
     \\
  &&\quad\quad\quad\quad
    \quad
     (\omega_{\hat{i}\hat{0}})_{\hat{j}}^{(r = 0)}\,\, 
  =\,\,-\,{\cal{B}}_{\hat{i} \hat{j}}^{\,(\rm g)}, 
\nonumber
\\
  &&\quad\quad\quad\quad
     \quad
     (\omega_{\hat{i} \hat{j}})_{\hat{k}}^{(r = 0)}\,\,=\,\,\,0.
\label{Ricci.connections.spat.local.frame}
\end{eqnarray}
\unboldmath
\end{itemize}
These LONB-connections $(\omega_{\hat{a} \hat{b}})_{\hat{c}}$
  in the frame of a {\it noninertial primary observer}
  are the key for proving our exact, explicit, and simple
  equations of motion,
  Eqs.~(\ref{eqs.motion.p.noninertial.2}),
  and our exact, explicit, and simple gravito-Maxwell equations,
  Sects.~\ref{gravito.Gauss.x}-\ref{div.grav.magn}.  

\subsection{Equations of motion in local frame
  \protect\\
 of GR-noninertial primary observer
\label{eqs.of.motion.noninert.obs.xxx}}

The equation of motion for the momentum~$\bar{p}$ of a
test-particle in gravitational and electromagnetic fields is,
\begin{eqnarray}
       \frac{d}{dt} p^{\hat{a}}   
      &=& -(\omega^{\hat{a}}_{\,\,\hat{b}})_{\hat{c}}\,
      p^{\hat{b}} \frac{dx^{\hat{c}}}{dt}
       + qF^{\hat{a}}_{\,\,\hat{b}}\frac{dx^{\hat{b}}}{dt}.
      \nonumber
\end{eqnarray}
In a general frame
the equation of motion has 24 Ricci connection
components~$(\omega^{\hat{a}}_{\,\,\hat{b}})_{\hat{c}}$
resp. 40 Christoffel-connection
components~$\Gamma^{\alpha}_{\,\,\beta\gamma}.$

But in our {\it frame} of LONBs
for a GR-{\it noninertial} primary {\it observer}
and his auxiliary observers,
the Ricci LONB-connections
$(\omega^{\,\hat{b}}_{\,\,\,\hat{a}})_{\hat{c}}$
in his time direction and his spatial directions are given 
by $(\vec{\cal{E}}_{\rm g}, \vec{\cal{B}}_{\rm g})$
in the very simple Eqs.~(\ref{Ricci.connections.spat.local.frame}).
This gives the exact GR-equations of motion
in our local reference frame of LONBs
for a {\it non-inertial} primary observer and
for particles starting on the primary worldline
in arbitrarily strong gravitational and electromagnetic fields:
exact, explicit, and strikingly simple, 
\begin{eqnarray}
\mbox{exact GR}&&\mbox{for relativistic test-particles}
\nonumber
\\ 
  (\frac{d}{d\hat{t}})_{\rm prim.obs}\,\,p^{\hat{i}}\,
   &=&\,   
       \varepsilon\, (\, \vec{\cal{E}}_{\rm g}
+ \,  2\vec{v} \times \vec{\cal{B}}_{\rm g} \, )^{\hat{i}} 
\,\,\,\,\,\mbox{gravit. forces} 
\nonumber
\\ 
&+&\, 
q\,(\,\vec{E}\,\, 
+\,\vec{v}\times\vec{B}\,)^{\hat{i}}, 
\,\,\,\,\mbox{el.mag. forces},   
\nonumber
  \\
(\frac{d}{d\hat{t}})_{\rm prim.obs} \,\,\varepsilon\, 
&=&\,  
\varepsilon \, \,(\vec{v} \cdot \vec{\cal{E}}_{\rm g})
\, \, \,+ \, \, \,  q \,  \,(\vec{v}\cdot\vec{E}),
\label{eqs.motion.p.noninertial.2}
\end{eqnarray}
where $\vec{v} \equiv (d\vec{x}/d\hat{t}\,)$
is the 3-velocity measured by the primary observer
     in his Minkowski tangent space
on his worldline.   
A particle's total relativistic energy measured by the observer
is $\varepsilon$,
the relativistic 3-momentum is
$\vec{p} = \varepsilon \, \vec{v}$,
for photons~$\varepsilon = \hbar \,\omega,$
and $(d/d\hat{t}\,)_{\rm prim.obs}$
refers to the time-difference
measured by the primary observer
on his chronometer.

In contrast to Eq.~(\ref{eqs.motion.p.noninertial.2}),
the {\it explicit} exact equations of motion in  {\it coordinate bases} 
are {\it complicated}.

Eqs.~(\ref{eqs.motion.p.noninertial.2})
are also the equations of motion for {\it Special Relativity}
in the local {\it frame} of a not-freefalling observer 
who is rotating relative spin-axes of gyros.  

For our exact GR-equations of motion,  
an {\it initial measurement} of a particle momentum $\bar{p}$
is made in the tangent space~$TM_P$ with~$P$ on the
{\it worldline} of the {\it primary observer}.
A {\it second measurement} is made by a neighbouring
observer (adapted to the primary observer)
at an infinitesimal time $\delta t$ later.  

The gravitational fields $(\vec{\cal{E}}_{\,\rm g},
\vec{\cal{B}}_{\,\rm g})$
are {\it not 3-vectors}.
The  gravitational fields ${\cal{F}}_{\hat{a}\hat{b}}^{(\rm g)}$
are identical with minus the Ricci LONB-connections 
$({\omega}_{\hat{a}\hat{b}})_{\hat{0}}.$

For spins, the exact equation of motion for
GR-{\it inertial} test-particles
(i.e. influenced only by gravitational fields)
in the {\it local frame}
of a {\it noninertial} primary observer is,
\begin{eqnarray}
  (\frac{d}{d\hat{t}})_{\rm prim.obs} S^{\hat{i}}
  &=&(\vec{S}\times\vec{\cal{B}}_{\rm g})^{\hat{i}}
      \nonumber
  \\
  &&+(\vec{S}\cdot\vec{v})(\vec{\cal{E}}_{\rm g}
      +\vec{v}\times\vec{\cal{B}}_{\rm g})^{\hat{i}},
\label{eq.motion.spin}
\end{eqnarray}
where we have used $(\bar{S}\cdot\bar{p}) = 0$, hence
$S^{\hat{0}} = (\vec{S}\cdot\vec{v}).$

\subsubsection{Fictitious Coriolis acceleration GR-equivalent
  \protect\\
  with contribution to
  gravito-magnetic acceleration
\label{equivalence.fict.gravit.forces}}

  Since GR has no Newton-inertial frames,
  the classical fictitious Coriolis acceleration
  (in a frame rotating relative to Newton-inertial),
\begin{eqnarray}
  \vec{a}_{\,\rm Coriolis}^{\,(\rm classical)} \,
     &=&\,
       2\,\vec{v}\times
\vec{\Omega}_{\,\rm frame\,of\,GR-obs}^{\,(\rm rel.to\,Newton-inertial)},
       \nonumber
\end{eqnarray}
  must be declared Einstein-{\it equivalent}
  with a contribution to the
  {\it gravitomagnetic acceleration}
  (relative to the GR-observer)
  {\it without matter-sources},
  \begin{eqnarray}
    \vec{\mathfrak{a}}_{\,\rm grav.magn}^{\,(\rm GR)}\,
          &=&\,2\,\vec{v}\times
          \vec{\cal{B}}_{\,\rm g}^{\,(\rm rel.to\,obs.LONB)},
              \nonumber
  \end{eqnarray}
  with $\vec{\cal{B}}_{\,\rm g}^{\,(\rm GR)}
  = -\,\vec{\Omega}_{\,\rm gyro}^{\,(\rm rel.to\,obs.LONB)}$:
the gravito-magnetic field 
  $\, \vec{\cal{B}}_{\rm g}$ of GR
  is generated in part by energy currents 
  and in part as a fictitious gravito-magnetic field. 

   \subsection{Spatially Local Coordinates 
     \protect\\
     of GR-noninertial observer
     \label{local.coord.noninertial.obs}}

For an inertial or noninertial primary observer
with his LONBs in the tangent spaces on his worldline,
we choose our {\it slicing} of spacetime
by hypersurfaces~$\Sigma_t$
for first-order radial separations from the primary observer:

$\Sigma_t$~starts Minkowski-{\it orthogonal}
to the primary worldline, and
our~$\Sigma_t$~is generated by {\it radial 4-geodesics}.

$\Sigma_t$~has as time-coordinate $t$  the time measured
on the wristwatch of the primary observer.

To first order in radial distance $r$
from the primary worldline,
the {\it intrinsic} geometry of~$\,\Sigma_t$ is {\it Euclidean},
and we choose {\it Cartesian} coordinates $x^i$ oriented
in the directions of the primary observer's
LONBs $\bar{e}_{\hat{i}}$.

The lines of fixed 3-coordinates $x^i$ are 
the worldlines of auxiliary observers adapted
to the primary observer.

The {\it lapse} function~$\alpha$
between slices $\Sigma_t$ is defined as the
{\it elapsed measured time}~$\delta \tau$
per increase in {\it coordinate-time}~$\delta t$
along the normal~$\bar{n}$
on the hypersurface~$\Sigma_t$, 
\begin{eqnarray}
\mbox{lapse}\,\,\equiv\,\,\alpha\, 
&\equiv&\,(\frac{d \tau}{dt})_{\bar{n} (\Sigma_t)}.
\nonumber
\end{eqnarray}

The {\it shift}~$\vec{\beta}$  is defined as 
shift of the {\it time coordinate line}
(= worldline of auxiliary observer)
from the {\it normal} on~$\Sigma_t$ 
per coordinate time~$t$, sign convention of MTW~\cite{MTW},
\begin{eqnarray}
  \mbox{shift}\,\,\equiv\,\,\beta^{\,i}\,  
  &\equiv&\,
  (v^i)_{\rm time-coord.line}^{({\rm rel.to}\,\bar{n}_{\Sigma})}\,\,
           =\,\,- \, (\frac{dx^i}{dt})_{\bar{n} (\Sigma_t)}.
           \nonumber
\end{eqnarray}

\subsubsection{Local coordinates of GR-inertial observer}
The spatially local coordinates
for an {\it inertial} primary observer
to first order in $\delta r$ have,
\begin{eqnarray}
  \mbox{lapse}\,\equiv\,\alpha\,=\,1,\,
     &\quad&\,
     \mbox{shift}\,\equiv\,\vec{\beta}\,=\,0.
     \nonumber 
  \end{eqnarray}

  In the local frame
  of an {\it inertial} primary observer
  with his adapted auxiliary observers
  and with his adapted spatially local coordinates,
  the crucial result for computing space-time curvature is
  that the primary and auxiliary observers have,
  \begin{eqnarray}
    &&\mbox{coordinate bases}\,\,\bar{e}_{a} \,\,=\,\,
       \mbox{LONBs}\,\,\bar{e}_{\hat{a}}
       \label{coord.bases.eq.LONBs}
  \end{eqnarray}

  On the {\it entire worldline} of a primary inertial observer,
  the metric is~$\eta_{\mu\nu}$,
  and the first derivatives of the metric
  in the {\it observer's time-} and {\it 3-directions} vanish,
  \begin{eqnarray}
    g_{\mu\nu}^{(r = 0)}\,=\,\eta_{\mu\nu},\,\,\,
    &&\,\,\,g_{\mu\nu,\lambda}^{(r = 0)}\,=\,0.
  \end{eqnarray}
  These coordinates are
  {\it new} and fundamentally {\it different} from the
  ``local inertial coordinates'' of GR-textbooks.

\subsubsection{Local coordinates of GR-noninertial observer}

We first consider a {\it nonspinning}
primary observer {\it accelerated} in the positive $x$-direction,
${\cal{E}}_{\rm g}^{x} = 
-\,({a}_{\rm obs}^{x})_{\rm rel.to\,ff}.$
In an infinitesimal time,
the accelerated observer's {\it time axis} at $(t_0 + \delta t)$
is {\it tilted} infinitesimally 
(relative to the observer's time axis at $t_0$) 
towards the direction of acceleration.
The accelerated observer's  $x$-{\it axis},
a line of {\it constant time},
is tilted towards the inertial positive $t$ axis
by the same positive amount
$(- {\cal{E}}^x_{\rm g}  \, \delta t)$,
\begin{eqnarray} 
\alpha_P \,
&=&\, 1 \,\, - \,\, 
\vec{\cal{E}}^{\,(\rm g)}_{P_0} \cdot \delta \vec{r}_P.
\nonumber
\end{eqnarray}
For a freefalling primary observer, rotating or nonrotating,
the lapse $\alpha$ is zero.
Hence the last equation is valid for any primary observer.

Now consider  a {\it spinning-freefalling} primary observer.
His adapted auxiliary observers are at time-independent separations,
hence at fixed values of our $r$-coordinate.
His auxiliary observers are {\it co-orbiting}  
with the 3-LONBs of the spinning primary observer,
hence at fixed $(\theta, \phi)$.
In our coordinates,
the {\it time-coordinate lines}
coincide with the {\it worldlines} of
our adapted {\it auxiliary observers}.

The shift vector $\vec{\beta}$
to first order  in $\, \delta \vec{r}_P \, $ 
is equal to the nonrelativistic velocity 
of the auxiliary observer $\vec{v}_{\,\rm aux.obs}$
relative to the normals on $\Sigma_t$
and equal to his nonrelativistic 
orbital velocity relative to the primary observer, 
\begin{eqnarray}
  \vec{\beta}^{\,(P)}
  &=& 
 (\vec{v}^{\,(P)})_{\,\rm aux.obs}^{\,(\rm rel.normal\, on\,\Sigma)}
      \nonumber
\\
  &=& \vec{\Omega}^{(P_0)}_{\rm prim.obs}
    \times  \delta \vec{r}^{\,(P)}_{\,\rm aux.obs}\,
    =\,-\,\vec{\cal{B}}_{\,\rm g}^{\,(P_0)} \times
    \delta \vec{r}^{\,(P)},
\nonumber
\\
\beta_i^{\, (P)} &=&   
({\cal{B}}_{ij}^{(g)})^{(P_0)} \,\, \delta r_{j}^{(P)}. 
\nonumber
\end{eqnarray}

For our spatially local coordinates
of a primary observer,
the shift is zero for a nonrotating primary observer,
freefalling or not freefalling.
Hence the last equation is valid
for any primary observer.

On the entire worldline of a chosen non-inertial observer
($r = 0),$
the metric $g_{\mu\nu}^{(r = 0)}$ and its first derivatives
in the observer's time direction are,
\begin{eqnarray}
  g_{\mu\nu}^{(r = 0)}\,\,
  =\,\,\eta_{\mu\nu},  
  &\quad\,\,\,&
  g_{\mu\nu,0}^{(r = 0)}\,\,
  =\,\,0.
      \nonumber
\end{eqnarray}

To first order in $\delta r$, the intrinsic geometry of $\Sigma_t$
is Euclidean, and we have chosen Cartesian 3-coordinates,
\begin{eqnarray}
g_{ij,k}^{(r = 0)}\,
  &=&\, 0.
      \nonumber
\end{eqnarray}

With $\,{\cal{E}}^{\hat{i}}_{\rm g} \equiv
-\,(a^{\hat{i}})_{\rm prim.obs.acceleration}^{(\rm rel.to\,GR-inertial)},$
an infinitesimal Lorentz boost of the primary observer
relative to GR-inertial
tilts his local time-axis and spatial axes
relative to GR-inertial.
The resulting lapse and shift give,  
\begin{eqnarray}
  g_{00}\,=\,-\,\alpha^2,
  \,\,\,&\quad&\,\,\, g_{0i}\,=\,\beta_i,
                \nonumber
  \\
   r = 0:\,\,\, g_{00,i}\,\,
  &=&\,\,\,2 \,{\cal{E}}_i^{(\rm g)},
\nonumber
\\
   g_{0i,j}\,\,
   &=&\,\, -\,{\cal{B}}^{(\rm g)}_{ij}.
              \nonumber
    \label{deriv.of.metric.E.B}                 
\end{eqnarray}

\section{Intrinsic Curvature
  \protect\\
with 
Cartan's LONB-method
\label{sec.curvature.Cartan}}

Cartan's LONB-method is seldom used in research papers
and not taught in most graduate
GR-courses, but we need it.~---
To introduce Cartan's LONB-method for computing 
curvature  we start in a 2-space. 
A closed infinitesimal curve~${\cal{C}}$, positively oriented,
has an integral of LONB-rotation
angles $(d\alpha/dx^{\mu}) = \omega_{\mu}$ (relative
to parallel transport),
\begin{eqnarray}
  \oint_{\cal{C}} \omega_{\mu}\,dx^{\mu}\,
  &\equiv&\,-\,(\delta \alpha)_{\cal{C}},
\end{eqnarray}
where $(\delta \alpha)_{\cal{C}}$ is the
{\it deficit} LONB-rotation angle.
The Gauss curvature $R_{\rm Gauss}$ is
$(\delta \alpha)_{\cal{C}}$
divided by the measured area $A_{\cal{C}}$
inside the boundary ${\cal{C}}$
for ${\cal{C}}\rightarrow 0$,
\begin{eqnarray}
  R_{\rm Gauss}\,
  &=&\,\lim_{{\cal{C}}\rightarrow 0}\, [\,A^{-1}_{\cal{C}}\,
      (\delta \alpha)_{\cal{C}}\,].
\end{eqnarray}

\subsection{Cartan's curvature equation
  \protect\\
  from roundtrip by parallel transport
\label{Cartan.curvature.eq.xxx}}

In curved (3+1) space, the  Riemann curvature tensor at $P$ is
defined  by the infinitesimal {\it deficit}
     LONB {\it Lorentz transformation}~$\,(\delta \,
   L^{\,\hat{a}}_{\,\, \, \, \hat{b}})_{\cal{C}}\,$
   after a round trip by parallel transport
   along an infinitesimal
   closed curve~${\cal{C}}$ around~$P$. 

   In parallel transport  of
   a vector,     
   its  LONB-components   
   transform, 
\begin{eqnarray}
\partial_{\nu}\,V^{\hat{a}}\,  
&=& \, - \,\,
    (\omega^{\hat{a}}_{\,\,\,\hat{b}})_{\nu}\,\, V^{\hat{b}},         
  \nonumber
\end{eqnarray}
where   $ (\omega^{\hat{a}}_{\, \, \, \hat{b}})_{\nu}$
is Cartan's LONB-connection with the
displacement coordinate-component~$\nu$.
We expand both factors around  an initial point~$P_0$ 
to first order in~$\delta x$,
 \begin{eqnarray}
  (\omega^{\hat{a}}_{\,\,\,\hat{b}})_{\nu}^{(x)}\, 
   &\approx&\,(\omega^{\hat{a}}_{\,\,\,\hat{b}})_{\nu}^{\,(x_0)}
    \,\,\, + \,\,\,
    [\,\partial_{\mu}\,(\omega^{\hat{a}}_{\,\,\, \hat{b}})_{\nu}]^{(x_0)}
    \,\,\,(x - x_0)^{\mu}.
       \nonumber
   \end{eqnarray}
   In parallel transport, the expansion of the
   vector field components~$V^{\hat{b}}$ gives,
   \begin{eqnarray}
    V^{\hat{b}}_{ (x)} \,
     &\approx& \,V^{\hat{b}}_{\, (x_0)} \,\,\, - \,\,\,
               [\,(\omega^{\hat{b}}_{\,\,\,\hat{c}})_{\mu}
               \,V^{\hat{c}}\,]_{ (x_0)} \,\,\,(x - x_0)^{\mu}.
         \nonumber
   \end{eqnarray}
   We insert the last two expansions into the  first equation
   and integrate along the closed curve  ${\cal{C}}.$

   The product of  constant terms gives
   $\oint d x^{\mu} = 0.$
   The products which are linear in $(x-x_0)$ give
   the {\it deficit Lorentz transformation}
   $\delta_{\cal{C}}L^{\hat{a}}_{\,  \, \hat{b}}$
   after an infinitesimal {\it roundtrip} by
     {\it parallel transport} around
   ${\cal{C}}$,
   \begin{eqnarray}
     &&\quad\,
        \delta_{\cal{C}}\,V^{\hat{a}}\,
        =\,\oint_{\cal{C}} dx^{\nu}\,\partial_{\nu} (V^{\hat{a}})\,
        =\,(\delta_{\cal{C}} L^{\hat{a}}_{\,\,\hat{b}})\,V^{\hat{b}}\,
        =
        \nonumber
     \\
     &=&  \frac{1}{2}\,\{ [\,\partial_{\mu}
        (\omega^{\hat{a}}_{\, \, \hat{b}})_{\nu} 
        +  (\omega^{\hat{a}}_{\, \,  \hat{s}})_{\mu} \, 
         (\omega^{\hat{s}}_{\, \,  \hat{b}})_{\nu}]
         - [\mu \Leftrightarrow \nu] \}_{(x_0)} \,
        V^{\hat{b}}_{(x_0)} \times
        \nonumber
     \\
     &&
        \quad \quad \quad
        \quad \quad \, \, 
        \times \oint_{\cal{C}}  (x - x_0)^{\mu}\,dx^{\nu}.
        \nonumber
        \label{deficit.Lorentz.trsf}
   \end{eqnarray}

   The {\it Riemann curvature 2-form}
   $({\cal{R}}^{\hat{a}}_{\, \, \, \hat{b}})_{\mu \nu}$
   is defined via the
   infinitesimal deficit Lorentz transformation
   $\delta_{\cal{C}}L^{\hat{a}}_{\,  \, \hat{b}}$
   after the roundtrip around an infinitesimal ${\cal{C}}$,
   \begin{eqnarray}
            \delta_{\cal{C}}L^{\hat{a}}_{\,  \, \hat{b}} \,
        &\equiv& \,
        \frac{1}{2}\, ({\cal{R}}^{\hat{a}}_{\, \, \, \hat{b}})_{\mu \nu} \,
        \oint_{\cal{C}} (x-x_0)^{\mu} dx^{\nu}.
     \label{op.def.Riemann.curv.2.form}
   \end{eqnarray}         
   The infinitesimal roundtrip integration
   in coordinate space $[\, \mu, \nu \,]$
   calls for the 2-form components $(...)_{\mu \nu}.$

The {\it antisymmetric derivative}
of a 1-form $\tilde{\sigma}$,
called {\it exterior derivative} and denoted by $d\tilde{\sigma}$,
produces a 2-form,
\begin{eqnarray}
&&d\,\tilde{\sigma}
  \,  \,\Leftrightarrow \,\, 
[\, d \, \tilde{\sigma} \, ]_{\mu \nu}  
\,\,\equiv\,\,     
\partial_{\mu} \,\sigma_{\nu}  - 
\partial_{\nu} \,\sigma_{\mu}. 
     \nonumber
\end{eqnarray}

   The {\it antisymmetric product} of two 1-forms, called
   {\it exterior product} and    
   {\it wedge product},
   produces a 2-form,
\begin{eqnarray}
          \tilde{\sigma} \wedge \tilde{\rho}
  \,\,\,\Leftrightarrow\,\,\, 
[\, \tilde{\sigma} \wedge \tilde{\rho}\, ]_{\mu \nu}
   \,\,\,&\equiv& \,\,\,
   \sigma_{\mu}\,\rho_{\nu} - \sigma_{\nu}\,\rho_{\mu}.
              \nonumber
\end{eqnarray} 

With the infinitesimal
deficit Lorentz transformation after an infinitesimal
roundtrip by parallel transport  
$\, \delta_{\cal{C}} L^{\hat{a}}_{\,  \, \hat{b}}\,$
and with the operational definition of the Riemann curvature 2-form
$\,({\cal{R}}^{\hat{a}}_{\, \, \, \hat{b}})_{\mu \nu}\,$
in Eq.~(\ref{op.def.Riemann.curv.2.form})
follows {\it Cartan's Riemann curvature 2-form}  equation,
\begin{eqnarray}
         (\,\tilde{\cal{R}}^{\hat{a}}_{\, \, \,  \hat{b}}\,)_{\mu \nu}\,
     &=&\,(\,d\, {\omega}^{\hat{a}}_{\, \, \, \hat{b}}
       \,\,  + \,\, \tilde{\omega}^{\hat{a}}_{\, \, \, \hat{s}}
         \wedge
         \tilde{\omega}^{\hat{s}}_{\, \, \, \hat{b}}\,)_{\mu \nu}.
         \label{Cartan.curvature.eq.x}
   \end{eqnarray}   
   Cartan's Riemann curvature 2-form 
   $(\tilde{{\cal{R}}}_{\hat{a} \hat{b}})_{\mu \nu}$
   with LONB-indices written in lower positions has: 

   (1) the  {\it antisymmetric} {\it LONB-index pair}
    $[\,\hat{a}, \, \hat{b}\,]$ 
    for the infinitesimal deficit  Lorentz transformation
    $\, \delta_{\cal{C}} L_{\hat{a} \hat{b}}\,$
    after a roundtrip by parallel transport,

    (2) the {\it antisymmetric coordinate-derivative pair}
    $  [\, \mu, \, \nu \,] $ 
    of the  closed displacement plaquette for the round trip. 

      The Riemann tensor and its two parts, the Ricci and Weyl tensors,
      do not depend on whether the chosen observer is GR-inertial
      or non-inertial.
    But Cartan's LONB-connections $(\omega^{\hat{a}}_{\,\,\hat{b}})_{\mu}$
    depend on the observer's acceleration/rotation relative to GR-inertial.

   \section{Gravito-Gauss law of GR
     \protect\\
     in frame of inertial observer
    \label{ff.accel.diff.inert.obs}}

   \subsection{Freefall acceleration difference
     \protect\\
     in local frame of inertial observer}

   Our crucial new tool, the 
   {\it frame} of an {\it inertial primary observer},
   includes his adapted auxiliary observers with their LONBs,
   Eq.~(\ref{conn.inert.obs}).
We now show that in this frame
$(R^{\,\hat{0}}_{\,\,\,\hat{1}})_{\hat{0}\hat{1}}
     \,\,=\,\,\partial_{\hat{1}}{\cal{E}}^{\,(\rm g)}_{\hat{1}}$.

We compute the {\it geodesic deviation},
the {\it acceleration-difference}
of freefalling particles infinitesimally separated,
initially at rest in the frame
of the chosen {\it inertial} primary observer.

This freefall acceleration-difference
in the 
direction~$\bar{e}_{\hat{1}}$ 
divided by~$(\delta x)^{\hat{1}}$   
is equal to~$(\partial_{\hat{1}}
{\mathfrak{g}}_{\,\rm GR}^{\hat{1}})_{\rm quasistat.part}   
\,\equiv\, \partial_{\hat{1}} {\cal{E}}^{\hat{1}}_{\rm g}.$

To compute the
curvature component~$(R^{\,\hat{0}}_{\,\,\,\hat{1}})_{01}$,
one needs the space-time
coordinate-plaquette~$[\,\partial_t,\partial_1].$~---
In the local coordinates adapted to a
GR-{\it inertial primary observer},
and going along the four sides of
the space-time plaquette~$[\,\partial_t,\partial_1]$
starting on his worldline,
the LONBs~$\bar{e}_{\hat{a}}$ and
his coordinate bases~$\bar{e}_{a} \equiv \partial_a$
are equal, 
    Eq.~(\ref{coord.bases.eq.LONBs}),
  \begin{eqnarray}
    &&\mbox{in local frame of GR-inertial primary observer:}
       \nonumber
       \\
    &&\quad\quad\quad\quad
      \quad\quad\quad\quad
       \bar{e}_{\hat{a}}\,\,=\,\,\bar{e}_{a}.
        \nonumber
        \end{eqnarray}

To compute curvature, one needs
LONB-connections $(\omega^{\,\hat{b}}_{\,\,\,\hat{a}})_{c}$
along the four sides of the coordinate plaquette.
Along the worldline
of the primary GR-inertial observer, 
the LONB-connection
vanishes, $(\omega^{\,\hat{b}}_{\,\,\,\hat{a}})_{\hat{0}}^{(r = 0)} = 0.$
For the two spatial displacements
starting on the primary worldline,
the LONB-connections also vanish
in the GR-inertial frame,
$(\omega^{\,\hat{b}}_{\,\,\,\hat{a}})_{\hat{i}}^{(r = 0)}\,=\,0$.
Along the 4th side of the coordinate plaquette,
along the worldline of the adapted auxiliary observer
(who is at fixed radial distance from the primary observer and
GR-noninertial because of tidal forces/torques),
$\,(\omega^{\,\hat{b}}_{\,\,\,\hat{a}})_{\hat{0}}^{(\rm aux.obs)}\neq\,0$.
Conclusion: 
\begin{itemize}
\item in Cartan's curvature equation
on the worldline of a GR-{\it inertial primary observer} 
in his  local frame 
with his adapted auxiliary observers and coordinates:
the {\it bilinear terms, wedge terms, vanish},
\begin{eqnarray}
  &&\quad\quad\quad\,\,\,\,\,
     (\tilde{\omega}^{\,\hat{b}}_{\,\,\,\hat{c}}\wedge
      \tilde{\omega}^{\,\hat{c}}_{\,\,\,\hat{a}})_{\mu\nu}\,
  =\, 0,
  \nonumber                                               
\\
  &&(R^{\,\hat{0}}_{\,\,\,\hat{1}})_{\hat{0}\hat{1}}
     \,\,
  =\,\,-\,\partial_{\hat{1}}(\omega^{\hat{0}}_{\,\,\,\hat{1}})_{\hat{0}}
     \,\,=\,\,\partial_{\hat{1}}{\cal{E}}^{\,(\rm g)}_{\hat{1}}.
     \label{freefall.acceleration.difference}
\end{eqnarray}
This {\it exact} equation
is {\it linear} in~$\vec{\cal{E}}_{\rm g}
= \vec{\mathfrak{g}}_{\,\rm quasistat}^{\,(\rm GR)}.$

In the frame of a GR-{\it inertial} observer,
the freefall acceleration
difference~$\partial_{\hat{1}}{\mathfrak{g}}
^{(\rm quasistat)}_{\hat{1}}$ is given by the Riemann tensor.
\item
But in the reference frame of a GR-{\it noninertial} observer,
the measured {\it freefall acceleration-difference}
is {\it not} given
by the Riemann tensor as demonstrated in
Sects.~\ref{gravito.Gauss.x} and
\ref{R00.not.rel.accel}.
\end{itemize}

\boldmath
\subsection{$ R^{\,\hat{0}}_{\,\,\,\hat{0}}
  =  \mbox{div}\,\vec{\cal{E}}_{\rm g}$
  in frame of  inertial observer
\label{R00.div.E}}
\unboldmath

In our local {\it frame} of LONBs for a GR-{\it inertial} observer
the exact explicit differential expression for  $R^0_{\,\,0}$
is extremely simple:
$R^0_{\,\,0}$ is identical
with $\partial_{\hat{i}} {\cal{E}}^{\hat{i}} =
\mbox{div}\,\vec{\cal{E}}_{\rm g}$.

In the {\it reference frame}
of a GR-{\it inertial} primary observer,
the {\it freefall acceleration-difference} (= geodesic deviation)
of a quasistatic particle and the primary inertial observer
(divided by their infinitesimal separation) is given by
Eq.~(\ref{freefall.acceleration.difference}).~---
The spherical average of the freefall-acceleration difference gives  
$\partial_{\hat{i}}{\mathfrak{g}}^{\hat{i}}_{\rm quasistat}
= \partial_{\hat{i}} {\cal{E}}_{\rm g}^{\hat{i}}
=\mbox{div}\,\vec{\cal{E}}_{\, \rm g}.$

In the local {\it reference frame} of a 
GR-{\it inertial} primary observer,
$\mbox{div}\,\vec{\cal{E}}_{\rm g}$ and
$R^{\,\hat{0}}_{\,\,\,\hat{0}}$
are given by the identical, exact, and
{\it linear differential expression}
in $\vec{\cal{E}}_{\, \rm g}
\equiv \vec{\mathfrak{g}}_{\,\rm GR}^{\,(\rm quasistat)}$,
\begin{eqnarray}
  &&\mbox{{\it reference frame}
     of GR-{\it inertial} primary observer:}
  \nonumber
  \\
  &&\quad\,\,\,\,
     R^{\,\hat{0}}_{\,\,\,\hat{0}}
     \,\,=\,\,\partial_{\hat{i}}{\cal{E}}_{\rm g}^{\hat{i}}
     \,\,=\,\,\mbox{div}\,
     \vec{\cal{E}}_{\rm g}^{\,(\rm GR-inert.obs.frame)},      
  \label{div.Ricci.0.0}
      \\
  &&\,\,\,\,\,
     \mbox{exact differential expression is {\it linear}  in}\,   
     {\cal{E}}^{\,(\rm g)}_{\hat{1}}.
\nonumber
\end{eqnarray}
This simple exact equation depends on
our new local reference frame of a GR-inertial observer,
Eq.~(\ref{conn.inert.obs}).

Later we shall prove the crucial fact:
in the reference frame of a GR-{\it non-inertial} observer,
$R^{\,\hat{0}}_{\,\,\,\hat{0}}
     \,\neq\,\mbox{div}\,\vec{\cal{E}}_{\rm g}^{\,(\rm GR)}$.

\subsection{From Newton-Gauss law to Einstein equations}
\label{Newton.Gauss.to Einstein}

Einstein's~$R^{\,\hat{0}}_{\,\,\,\hat{0}}$ equation
    for nonrelativistic matter and neglecting electromagnetic fields is 
     $R^{\,\hat{0}}_{\,\,\,\hat{0}}
       = - 4 \pi G \,\rho_{\rm mass}$. Our important result:
\begin{itemize}
\item In our {\it reference frame} of LONBs
  for a GR-{\it inertial observer},
for nonrelativistic physics
and neglecting electromagnetic fields,
       the {\it partial differential equation}
         for $R^{\,\hat{0}}_{\,\,\,\hat{0}}$
 is {\it identical} with the {\it Gauss law} for {\it Newton-gravity},
\begin{eqnarray}
R^{\hat{0}}_{\,\,\hat{0}}
      =\mbox{div}\,\vec{\mathfrak{g}}_{\rm GR}^{\,(\rm quasistat)} 
     &=&\mbox{div}\,\vec{g}_{\rm Newton}
      =-4\pi G\rho_{\rm m}.
\label{Newton.Gauss.Einstein}
\end{eqnarray}
Nonrelativistically $R^{\hat{0}}_{\,\,\hat{0}}= -4\pi G\rho_{\rm m}$
holds in the frame of any observer, GR-inertial and noninertial.
\item 
  From the {\it Gauss-Newton law}
  and the concept of timelike GR-geodesics (freefalling) follows
  Einstein's $R^{\,\hat{0}}_{\,\,\,\hat{0}}$ equation
  for nonrelativistic matter, Eq.~(\ref{Newton.Gauss.Einstein}).
  From this combined with Special Relativity and the contracted
  Bianchi identity 
  follow (as explained in GR-texts)
the full {\it Einstein equations},
\begin{eqnarray}
   R^{\,\hat{a} \hat{b}} 
  &=& 8\pi G\,(T^{\,\hat{a} \hat{b}}
      - \frac{1}{2}\eta^{\,\hat{a}\hat{b}} T),
      \nonumber
\end{eqnarray}
where $T^{\,\hat{a} \hat{b}} = $ energy-momentum
tensor of matter and electromagnetic fields,
and $T$ is its trace.
\end{itemize}
Einstein's $R^{\,\hat{0}}_{\,\,\,\hat{0}}$ equation
for relativistic sources follows,
\begin{eqnarray}
R^{\,\hat{0}}_{\,\,\,\hat{0}}\,
  &=&\, - 4 \pi G (\rho_{\varepsilon} + 3 \tilde{p})_{\rm matter+EM},
      \label{R00.eq.rel.sources}
      \end{eqnarray}
      where  $3 \tilde{p} \equiv$ trace of 3-{\it momentum-flow tensor}
      $\tilde{p}^{\,\hat{i}\hat{j}}.$~---
The 3-momentum-flow tensor 
  $\tilde{p}^{\,\hat{i}\hat{j}}$  
must be distinguished
from the {\it pressure} tensor  $p^{\,\hat{i}\hat{j}}$, 
which is measured in the
instantaneous rest-frame of a fluid element.~---
{\it Maxwell's stress} tensor
has the {\it opposite sign}
from the 3-momentum-flow tensor $\tilde{p}^{\,\hat{i}\hat{j}}$
of electromagnetism.~--- 
In Einstein's equation for $R^{\,\hat{0}}_{\,\,\,\hat{0}}$
the source from electromagnetic fields is, 
    \begin{eqnarray}
      (\rho_{\varepsilon} + 3\, \tilde{p})^{(\rm EM)}\,
      &=&\, 2\, \rho_{\varepsilon}^{(\rm EM)}\,
          =\, \frac{1}{4\pi} (\vec{E}^2 + \vec{B}^2).
          \label{EMsource.in.R00}
    \end{eqnarray}

\section{Gravito-Maxwell equations 
  \protect\\
  in frame of noninertial observer
\label{gravito.Maxwell.x}}

\boldmath
\subsection{Golden Rule for 
  Riemann tensor
 \protect\\
  in LONB-components
 $(R^{\hat{a}}_{\, \, \, \hat{b}})_{\hat{c} \hat{d}}$
\protect\\
from Ricci LONB-connections
 $(\omega^{\hat{a}}_{\, \, \, \hat{b}})_{\hat{c}}$
\label{sect.golden.rule}}
\unboldmath

  To obtain the Riemann tensor in LONB-components
  $(R^{\,\hat{a}}_{\,\,\,\hat{b}})_{\hat{c}\hat{d}}$
  in our 
  frame of LONBs for a non-inertial primary observer,
we start with Cartan's curvature equation,
\begin{eqnarray}
(\tilde{\cal{R}}^{\hat{a}}_{\, \, \, \hat{b}})_{\gamma\delta}\,
&=&\, (d\, \tilde{\omega}^{\hat{a}}_{\, \,  \, \hat{b}}
\,+\, \tilde{\omega}^{\hat{a}}_{\, \,  \, \hat{s}} \wedge
    \tilde{\omega}^{\hat{s}}_{\, \,  \, \hat{b}})_{\gamma\delta},
    \nonumber
\end{eqnarray}
where the pair of displacement subscripts  $\gamma\delta$
denotes the antisymmetric pair  
 $(\gamma\delta - \delta\gamma)$ in the coordinate basis, 
\begin{eqnarray}
  (R^{\,\hat{a}}_{\, \, \, \hat{b}})_{\gamma\delta}\,
&=&\,
[\,\partial_{\gamma} \, (\omega^{\hat{a}}_{\, \, \, \hat{b}})_{\delta} 
\,+\,  
(\omega^{\hat{a}}_{\, \, \, \hat{s}})_{\gamma}\, 
     (\omega^{\hat{s}}_{\,\,\,\hat{b}})_{\delta}
     \,  - \,
 [ \, \gamma \Leftrightarrow \delta  \, ].
\nonumber
\end{eqnarray}

But we need {\it all four indices} of the {\it same type}: 
(1)~for contracting the Riemann tensor to obtain the Ricci tensor,
(2)~for the 1st Bianchi identity.~---
We want directly {\it measurable} components,
hence LONB components.

We work in our frame of a non-inertial observer with the
LONB-connections in his time-direction and his 3-directions,
Eqs.~(\ref{Ricci.connections.spat.local.frame}).

To obtain the Riemann tensor
$(R^{\hat{a}}_{\, \, \,  \hat{b}})_{\hat{c} \hat{d}}^{\,(P)}$
in LONB-components,
we choose the coordinate basis $\bar{e}_{\gamma}^{\, (P)}$
equal to the LONB    $\,\bar{e}_{\hat{c}}^{\, (P)}$
at points $P$ on the {\it worldline} of the {\it primary observer},
\begin{eqnarray}
  \mbox{LONB at}\,\,P\,\,\equiv\,\,\bar{e}_{\hat{c}}^{\, (P)}
  \,&=&\,
  \bar{e}_{\gamma}^{\,(P)}\,\,\equiv\,\,\mbox{coord. basis at}\,\,P,
           \nonumber
  \\
  (R^{\hat{a}}_{\, \, \, \hat{b}})_{\hat{c} \hat{d}}^{(P)}
  \, \, &=& \, \,
            (R^{\hat{a}}_{\, \, \, \hat{b}})_{\gamma\delta}^{(P)}.
  \nonumber
\end{eqnarray}

We want to re-express Cartan's LONB-connections
for coordinate displacements
$(\omega^{\hat{a}}_{\,\,\, \hat{b}})_{\gamma}$
by Ricci's LONB-connections for LONB displacements 
$(\omega^{\hat{a}}_{\,\,\,\hat{b}})_{\hat{c}},$
\begin{eqnarray}
  (\omega^{\hat{a}}_{\,\,\,\hat{b}})_{\gamma}
  \,\,&=&\,\,
  (\omega^{\hat{a}}_{\,\,\,\hat{b}})_{\hat{s}}\,\,(e^{\hat{s}})_{\gamma}.
            \nonumber
\end{eqnarray}

In Cartan's curvature equation
we first consider the {\it wedge term}
$(\tilde{\omega} \wedge \tilde{\omega}),$
where all displacement vectors are in the tangent space at $P$
on the primary worldline,
hence the LONB-displacements are identical
with the coordinate-basis displacements.
This gives
Cartan's wedge term in terms of LONB-displacements, 
\begin{eqnarray}
(R^{\hat{a}}_{\,\,\,\hat{b}})_{\hat{c} \hat{d}}^{(
  \tilde{\omega}\wedge\tilde{\omega})}\,
   &=&\, 
 [ (\omega^{\hat{a}}_{\,\,\,\hat{s}})_{\hat{c}}\, 
   (\omega^{\hat{s}}_{\,\,\,\hat{b}})_{\hat{d}} ]
        -  [  \hat{c} \Leftrightarrow \hat{d}\, ].  
       \label{Cartan.wedge}
\end{eqnarray}

In Cartan's~{\it derivative term}~$(d\tilde{\omega})_{\gamma\delta},$
rewritten with the Ricci connection
$  (\omega^{\hat{a}}_{\, \, \, \hat{b}})_{\hat{c}},  $
the derivative $\partial_{\gamma}$ acts on the {\it product},
$\partial_{\gamma} \,
[(\omega^{\hat{a}}_{\,\,\,\hat{b}})_{\hat{s}}\,(e^{\hat{s}})_{\delta}].$

For the derivative $\partial_{\gamma}$
acting on the first factor,
the second factor is evaluated on the primary worldline, 
\begin{eqnarray}
  (e^{\hat{s}})_{\delta}^{(P)}  \,
  &=& \,  \delta^{\hat{s}}_{\delta}.
  \nonumber
\end{eqnarray}
The derivative $\partial_{\gamma}$ acting on the first factor
at $P$ on the primary worldline
is equal to the derivative $\partial_{\hat{c}}$,
\begin{eqnarray}
  (e^{\hat{s}})_{\delta}^{(P)}\,\,\,\partial_{\gamma}^{(P)}\,
  (\omega^{\hat{a}}_{\,\,\,\hat{b}})_{\hat{s}}\,\,
  &=&\,\, 
  \partial_{\hat{c}}\,(\omega^{\hat{a}}_{\,\,\,\hat{b}})_{\hat{d}},   
     \nonumber
\end{eqnarray}
This gives the result for
the LONB-derivative $\partial_{\hat{c}}$
acting on the Ricci LONB-connection
$(\omega^{\hat{a}}_{\, \, \, \hat{b}})_{\hat{d}}$,
\begin{eqnarray}
  (R^{\hat{a}}_{\, \, \,  \hat{b}})_{\hat{c} \hat{d}}^{ (  
  \partial    \tilde{\omega} ) }
  \, &=& \,  
 [ \,  \partial_{\hat{c}} \, 
   (\omega^{\hat{a}}_{\, \, \, \hat{b}})_{\hat{d}} \, ]
         \, - \, [ \, \hat{c} \Leftrightarrow \hat{d}  \, ].
         \label{Cartan.partial.omega}
\end{eqnarray}

For the  derivative $\partial_{\gamma}$
at $P$ on the primary worldline
acting on the  second  factor, 
$\partial_{\gamma}^{(P)}  \, (e^{\hat{s}})_{\delta},$
and after antisymmetrization $[\gamma \Leftrightarrow \delta]$,
we use Cartan's implicit equation for the LONB-connection
$\,\tilde{\omega}^{\, \hat{s}}_{\, \, \, \, \hat{r}}$,
\begin{eqnarray}
  d\tilde{e}^{\,\hat{s}}\,
  &=&\,  
     - \, \tilde{\omega}^{\, \hat{s}}_{\, \, \, \, \hat{r}}
     \wedge \tilde{e}^{\,\hat{r}}.
        \nonumber
\end{eqnarray}
With displacement indices in the coordinate-basis,
\begin{eqnarray}
  \partial_{\gamma} (e^{\hat{s}})_{\delta}
  - \partial_{\delta}\, (e^{\hat{s}})_{\gamma}\,
   &=&\,-\,(\omega^{\hat{s}}_{\,\,\,\hat{r}})_{\gamma} 
     \,(e^{\hat{r}})_{\delta}
     \,+\, 
     (\omega^{\hat{s}}_{\,\,\,\hat{r}})_{\delta} 
     \,(e^{\hat{r}})_{\gamma}.
           \nonumber
\end{eqnarray}
On the primary worldline,
$(e^{\hat{r}})_{\delta}
=\delta^{\,\hat{r}}_{\,\hat{d}}\,$
and
$\,\partial_{\gamma} = \partial_{\hat{c}},$
\begin{eqnarray}
 \partial_{\gamma} \, (e^{\hat{s}})_{\delta} \,- \,
   \partial_{\delta} \, (e^{\hat{s}})_{\gamma}  
  \, \, &=& \, 
            [ \, (\omega^{\hat{s}}_{\, \, \, \hat{c}})_{\hat{d}} \, ]
            - [ \, \hat{c}   \Leftrightarrow \hat{d} \, ].
            \nonumber
\end{eqnarray}
This gives the result for
the  LONB-derivative $\partial_{\hat{c}}$
at $P$ on the primary worldline
acting on $(e^{\hat{s}})_{\delta}$,
\begin{eqnarray}
  (R^{\,\hat{a}}_{\,\,\,\hat{b}})_{\hat{c}\hat{d}}
  ^{(\partial\tilde{e}^{\hat{s}})}
  \,&=&\,\,(\omega^{\hat{a}}_{\,\,\,\hat{b}})_{\hat{s}}\,\,
            [\,(\omega^{\hat{s}}_{\,\,\,\hat{c}})_{\hat{d}}
         - (\hat{c} \Leftrightarrow \hat{d})\, ].
\label{Cartan.d.e}
\end{eqnarray}

The three terms for
$(R^{\hat{a}}_{\, \, \,  \hat{b}})_{\hat{c} \hat{d}}$ 
from Eqs.~(\ref{Cartan.wedge}-\ref{Cartan.d.e}) added
give our Golden Rule
for the Riemann tensor in LONB-components
in our frame of a non-inertial primary
observer with the
Ricci LONB-connections $(\omega^{\hat{b}}_{\,\,\hat{a}})_{\hat{c}},$
\boldmath
\begin{eqnarray}
  &&\quad\quad\quad\quad
     \quad\quad\,\,\,\,
     \mbox{\bf Golden Rule}
     \nonumber
  \\
  &&\quad\quad\quad\quad\quad\quad\,\,\,
    \quad\,
     (R^{\,\hat{a}}_{\,\,\,\hat{b}})_{\hat{c}\hat{d}}\,=
 \nonumber
 \\
  &&  [ \,  \partial_{\hat{c}} \, 
      (\omega^{\hat{a}}_{\, \, \, \hat{b}})_{\hat{d}}
   \, + \,
        (\omega^{\hat{a}}_{\, \, \, \hat{s}})_{\hat{c}} \, 
        (\omega^{\hat{s}}_{\, \, \, \hat{b}})_{\hat{d}}
   \, + \,
        (\omega^{\hat{a}}_{\, \, \, \hat{b}})_{\hat{s}} \,
    (\omega^{\hat{s}}_{\, \, \, \hat{c}})_{\hat{d}} \, ]
    \nonumber
  \\
  &&\quad\quad\quad\quad
    \quad\quad\,\,\,\,
     - \, [ \,\hat{c}   \Leftrightarrow \hat{d} \, ].
     \label{golden.rule.curvature}
\end{eqnarray}
\unboldmath

\subsection
{Gravito-Gauss law of General Relativity
\label{gravito.Gauss.x}}

We derive the gravito-Gauss law of GR
in the observer's {\it reference frame} of LONBs with
the LONB-connections $(\omega^{\hat{b}}_{\,\,\hat{a}})_{\hat{c}}$.
We use:

(1)~the {\it Golden Rule},
Eq.~(\ref{golden.rule.curvature}),
for the Riemann tensor in
LONB components~$(R^{\,\hat{a}}_{\,\,\,\hat{b}})_{\hat{c}\hat{d}}$
expressed by Ricci
LONB-connections~$(\omega^{\hat{a}}_{\,\,\,\hat{b}})_{\hat{c}}$,

(2)~our local {\it frame} of an {\it observer}
with his adapted auxiliary observers
and his LONB-connections in his time-direction and 3-directions,
$(\omega^{\hat{b}}_{\,\,\,\hat{a}})_{\hat{c}},$
Eqs.~(\ref{Ricci.connections.spat.local.frame}).

First consider the {\it wedge term}
in Cartan's curvature equation 
which produces 
$(R^{\,\hat{0}}_{\,\,\,\hat{0}})^{(\tilde{\omega}\wedge\tilde{\omega})}$,
   Eq.~(\ref{Cartan.wedge}),
\begin{eqnarray}
&&(R^{\,\hat{0}}_{\,\,\,\hat{0}})^{(\tilde{\omega}\wedge\tilde{\omega})}
   \,\,=\,\,
   (R^{\,\hat{0}}_{\,\,\,\hat{i}}
   )_{\hat{0}\hat{i}}^{(\tilde{\omega}\wedge\tilde{\omega})}
\,\,\equiv\,\,
( \tilde{\omega}^{\hat{0}}_{\, \,  \, \hat{k}} \wedge
   \tilde{\omega}^{\hat{k}}_{\, \,  \, \hat{i}})_{\hat{0}  \hat{i}}
   \nonumber
  \\
  && \quad\quad\,\,=\,\,
  (\omega_{\hat{k}\hat{0}})_{\hat{0}}\,(\omega_{\hat{k}\hat{i}})_{\hat{i}}
  \,-\,
  (\omega_{\hat{k} \hat{0}})_{\hat{i}}\,(\omega_{\hat{k}\hat{i}})_{\hat{0}}
     \nonumber
  \\
  &&\quad\quad\quad\,\,\,\,\,
     =\,\,-\,{\cal{B}}_{\hat{k}\hat{i}}^{(\rm g)} \,
     {\cal{B}}_{\hat{k}\hat{i}}^{(\rm g)}
  \,\,=\,\,-\,2\,\vec{\cal{B}}_{\rm g}^{\,2},
  \nonumber
\end{eqnarray}
where the Ricci LONB-connections
$(\omega_{\hat{a}\hat{b}})_{\hat{c}}$
for our frame of a primary observer are from
Eq.~(\ref{Ricci.connections.spat.local.frame}).

Next consider the term in the Golden Rule where the
LONB derivative~$\partial_{\hat{c}}$
acts on the Ricci
LONB connection,
Eq.~(\ref{Cartan.partial.omega}),   
and use the Ricci
LONB connections~$(\omega_{\hat{a}\hat{b}})_{\hat{d}}$
in the {\it frame} of the primary {\it observer},
Eq.~(\ref{Ricci.connections.spat.local.frame}),
    \begin{eqnarray}
         (R^{\,\hat{0}}_{\, \, \, \hat{0}})^{(\partial\tilde{\omega})}  
       \,\,&=&\,\,
   (R^{\,\hat{0}}_{\,\,\,\hat{i}})_{\hat{0}\hat{i}}^{(\partial\tilde{\omega})}
     \nonumber
    \\
     =\,\, -\,\partial_{\hat{i}}\, 
       (\omega^{\hat{0}}_{\,\,\,\hat{i}})_{\hat{0}}\,\,
       &=&\,\,\partial_{\hat{i}} {\cal{E}}_{\rm g}^{\hat{i}}\,\,
      =\,\,\mbox{div}\,\vec{\cal{E}}_{\rm g}.
         \nonumber
\end{eqnarray}

Finally consider the term in the Golden Rule where the
LONB-derivative~$\partial_{\hat{c}}^{(P)}$
acts on $(\tilde{e}^{\hat{s}})_{\delta},$
Eq.~(\ref{Cartan.d.e}),
\begin{eqnarray}
  &&\quad\,\,
     (R^{\hat{0}}_{\,\,\,\hat{0}})^{(\partial\tilde{e}^{\hat{s}})}\,\,
  =\,\,
     (R^{\hat{0}}_{\,\,\,\hat{i}})_{\hat{0}\hat{i}}
     ^{(\partial\tilde{e}^{\hat{s}})}
     \,\,\equiv
      \nonumber
  \\
  &&\equiv\,\, (\omega^{\hat{0}}_{\, \, \, \hat{i}})_{\hat{s}} \,
            [ \, (\omega^{\hat{s}}_{\, \, \, \hat{0}})_{\hat{i}}
         - (\hat{0}   \Leftrightarrow \hat{i}) \, ]\,\,   
      =\,\,-\,\vec{\cal{E}}^{\,2}_{\rm g}.
       \nonumber
\end{eqnarray}

The sum of the three terms gives the exact 
explicit {\it differential expression}
for~$R^{\,\hat{0}}_{\, \, \, \hat{0}}$
in our {\it reference frame} of  any {\it primary observer}
with his LONB-connections,
  \begin{eqnarray}
    R^{\,\hat{0}}_{\, \, \, \hat{0}} \,
    &=&\,\mbox{div}\,\vec{\cal{E}}_{\rm g}
         \, - \, (\vec{\cal{E}}^{\,2}_{\rm g}
        \, + \, 2\,\vec{\cal{B}}^{\,2}_{\rm g}),
        \label{R00.div.E.g}
  \end{eqnarray}
where $\mbox{div}\,\vec{\cal{E}}_{\rm g} =
\partial_{\hat{i}} {\cal{E}}_{\rm g}^{\hat{i}}$
in our frame of LONBs.
  The simplicity of Eq.~(\ref{R00.div.E.g}) depends 
  on our reference frame of an observer
  with his LONB connections of
  Eqs.~(\ref{Ricci.connections.spat.local.frame}).
  But in general coordinates,
  the explicit expression for $R^0_{\,\,0}$ 
  is very complicated .~--- 
  The {\it structure} of the right-hand side
  in Eq.~(\ref{R00.div.E.g})
  (without numerical prefactors and signs)
  follows from the structure of our Golden Rule and from $J^P$:
  the derivative term must be linear in
  $\mbox{div}\,\vec{\cal{E}}_{\rm g}$,
  the bilinear terms must be proportional to 
  $\vec{\cal{E}}_{\rm g}^{\,2}$ and to $\vec{\cal{B}}_{\rm g}^{\,2}$.

The {\it acceleration-difference} 
of neighbouring freefalling particles,
spherically averaged, is given by 
$ \mbox{div}\,\vec{\cal{E}}_{\rm g}$.
   Our exact Eq.~(\ref{R00.div.E.g})
  shows the crucial {\it difference} between
  $R^{\,\hat{0}}_{\,\,\,\hat{0}}$
  and the {\it geodesic deviations} given
  by~$\mbox{div}\,\vec{\cal{E}}_{\rm g}$.
  The geodesic deviations depend on whether the observer
  (with his frame of LONBs) 
  is accelerated and/or rotating.~---
  Eq.~(\ref{R00.div.E.g}) contradicts conclusions in
  GR-texts~\cite{Wald, Weinberg, Poisson.Will, Hartle}
  that ``geodesic deviations are given by the Riemann tensor''.

The $R^{\,\hat{0}}_{\, \, \,  \hat{0}}$
equation of General Relativity states,
\begin{eqnarray}
  R^{\,\hat{0}}_{\,\,\,\hat{0}}
  \,&=&\,-\,4\pi G\, (\rho_{\varepsilon} + 3\tilde{p})_{\rm matter+EM},
        \nonumber
\end{eqnarray}
where $3\tilde{p}$ is the trace of the 3-momentum-flow tensor,
and $-4\pi G\, (\rho_{\varepsilon} + 3\tilde{p})_{\rm EM}  =
- G (\vec{E}^2 + \vec{B}^2)$ is the 
electromagnetic source 
for $R^{\,\hat{0}}_{\, \, \,  \hat{0}}$,
Eq.~(\ref{EMsource.in.R00}).

We combine the last two equations
to eliminate $R^{\,\hat{0}}_{\, \, \,  \hat{0}}$,
\boldmath
\begin{eqnarray}  
  &&\mbox{\bf gravito-Gauss law of General Relativity}
     \nonumber
  \\
  &&\mbox{\bf in 
     LONB-frame of noninertial observer}
     \nonumber
  \\
    &&\mbox{\bf div}\,\vec{\cal{E}}_{\rm g}
        =-4\pi G(\rho_{\varepsilon}+3\tilde{p})_{\rm matter}
  \nonumber
  \\
  &&\quad\quad\quad\quad\,
     -G(\vec{E}^2+\vec{B}^2)_{\rm EM}
     +(\vec{\cal{E}}_{\rm g}^{\,2}
     +2\vec{\cal{B}}_{\rm g}^{\,2}).
     \label{gravito.Gauss}
\end{eqnarray}
\unboldmath
$\mbox{div}\,\vec{\cal{E}}_{\rm g}$ and
$(\vec{\cal{E}}_{\rm g}^{\,2}+2\vec{\cal{B}}_{\rm g}^{\,2})$
depend on the observer (inertial versus non-inertial)
and his frame of LONBs.~---
Our gravito-Gauss law of GR  is exact, totally explicit,
simple, and new.
The simplicity of our exact gravito-Gauss law
is due to our reference frame
of LONBs for a GR-noninertial observer,
Eq.~(\ref{Ricci.connections.spat.local.frame}),
which is entirely new.

The  sources of the GR-gravitoelectric
field~$\vec{\cal{E}}_{\rm g}$ in the gravito-Gauss law
are contributed by {\it all sources},
including the gravitational sources, 
\boldmath
\begin{eqnarray}
  \mbox{\bf div}\vec{\cal{E}}_{\rm g}
  &=&-4\pi
      G(\rho_\varepsilon+3\tilde{p})_{\rm matter+EM+gravity}.
     \label{sources.div.E}
\end{eqnarray}
\unboldmath
In contrast, the sources in
Einstein's $R^{\,\hat{0}}_{\,\,\,\hat{0}}$ equation
do not include the gravitational sources,
\begin{eqnarray}
  R^{\,\hat{0}}_{\, \, \, \hat{0}} 
  &=&-4\pi G\,(\rho_{\varepsilon} + 3 \tilde{p})_{\rm matter+EM}.
      \nonumber
\end{eqnarray}

The {\it acceleration-difference} 
of neighbouring freefalling particles,
spherically averaged, is given by 
   $\mbox{div}\,\vec{\mathfrak{g}}_{\rm GR}
   \equiv \mbox{div}\,\vec{\cal{E}}_{\rm g}$
   and determined by
   the gravito-Gauss law of GR.~---
But $R^{\,\hat{0}}_{\, \, \, \hat{0}}$ 
does not determine the acceleration-difference of
freefalling particles in the frame of a noninertial observer.

\begin{itemize}
  \item
The sources of $\mbox{div}\,\vec{\cal{E}}_{\rm g}$
include the gravitational bilinears
$(\vec{\cal{E}}_{\rm g}^{\, 2}  
+ 2\vec{\cal{B}}_{\rm g}^{\, 2})$.
This  is a {\it dark source} for
{\it repulsive gravity}
in contrast to the
gravitationally attractive matter
and electromagnetic sources.
\end{itemize}

For the new terms $\vec{\cal{E}}_{\rm g}^{\, 2}$ and  
$ 2\vec{\cal{B}}_{\rm g}^{\, 2}$
we give {\it independent} and {\it elementary} derivations in
Sect.~\ref{R00.not.rel.accel},
which are based only on Einstein's concepts of 1911,
but not on Einstein's equations of 1915.

In the solar system, 
Fiducial Observers at fixed coordinates $(r, \theta, \phi)$
measure $\vec{\cal{E}}_{\rm g} \neq 0,$
hence  they measure {\it repulsive gravity} 
in $\mbox{div}\,\vec{\cal{E}}_{\rm g}$ in the halo of the Sun.

In our inhomogeneous-anisotropic universe,
the Fiducial Observers adapted to any of the standard coordinates
are {\it not} GR-{\it inertial},
and in our late universe the gravito-magnetic field is negligible in the
gravito-Gauss equation:
\begin{itemize}
  \item  {\it repulsive gravity} from 
    $(\rho_{\varepsilon}+3\tilde{p})_{\rm grav}
\approx -\vec{\cal{E}}_{\rm g}^{\,2}/(4\pi G)$ contributes to the
observed {\it accelerated expansion} of our {\it universe} today.
An important task is to determine the magnitude of this effect.
\end{itemize}

In the gravito-Maxwell laws of GR,
the  form of the bilinear expressions
in~$(\vec{\cal{E}}_{\rm g}, \vec{\cal{B}}_{\rm g})$
follows from covariance under~$J^P$, i.e.
under rotations and space-reflections:
the sources of $\mbox{div}\,\vec{\cal{E}}_{\rm g}$ have $J^P = 0^+$.

In the frame of a GR-{\it inertial} primary observer,
$\, R^{\,\hat{0}}_{\,\,\,\hat{0}}
= \mbox{div} \, \vec{\cal{E}}_{\rm g}$.
But in the frame of a GR-{\it noninertial} primary observer,
these two quantities are unequal:
for $\, R^{\,\hat{0}}_{\,\,\,\hat{0}}$ it is irrelevant, whether the
primary observer is {\it inertial} or {\it non-inertial}.
But for $\, \mbox{div} \, \vec{\cal{E}}_{\rm g}  \,$, 
all depends on whether
the primary observer is inertial versus non-inertial.

\subsubsection{Gravitational $(\rho_{\varepsilon} + 3 \tilde{p})$
depends on observer's  local frame}

Sources of $\mbox{div}\,\vec{\cal{E}}_{\rm g}$
in the gravito-Gauss law are
$(\rho_{\varepsilon} + 3 \tilde{p})$ of
(1)~matter, (2)~electromagnetic fields,
\begin{eqnarray}
  (\rho_{\varepsilon} + 3 \tilde{p})_{\rm EM} \,
  &=&\,(\vec{E}^2 + \vec{B}^2)/(4\pi),
  \nonumber
\end{eqnarray}
plus (3) a term which must be identified with
$(\rho_{\varepsilon} + 3 \tilde{p})$
of gravitational fields,  
\begin{eqnarray}
  (\rho_{\varepsilon} + 3 \tilde{p})_{\rm grav}\,
  &=&\,- (\vec{\cal{E}}_{\rm g}^{\,2}
      + 2 \vec{\cal{B}}^{\,2}_{\rm g})/(4\pi G).
  \label{grav.energy.plus.3p}
\end{eqnarray}

The gravitational $(\rho_{\varepsilon} + 3\tilde{p})_{P}$
depends on the {\it observer} (noninertial vs inertial)
with worldline through $P$
and his {\it frame} with 
connections~$(\omega^{\hat{b}}_{\,\,\,\hat{a}})_{\hat{c}}^{(P)}$.
In the frame of a 
{\it inertial} observer
with worldline through $P$,
$(\rho_{\varepsilon} + 3\tilde{p})_{\rm grav}^{(P)} = 0.$

Landau and Lifshitz derived the density of
gravitational energy, momentum, and momentum
flow~\cite{Landau.Lifshitz.class.fields.1951}.
Explicit expressions are also given in
MTW~\cite{MTW} and Weinberg~\cite{Weinberg}.
These authors worked with general coordinates, {\it not adapted}
to a chosen local {\it observer} with his {\it local frame}.
For a (3+1)-split their results give thousands of terms  
instead of our simple
results,~Eqs.~(\ref{grav.energy.plus.3p}, \ref{grav.energy.current}).

\subsubsection{Repulsive gravity: irrelevant in FLRW universe,
  \protect\\
but inescapable in strongly perturbed universe}

In a Friedmann-Lemaitre-Robertson-Walker universe
{\it natural observers} are comoving and nonrotating
relative to matter and electromagnetic energy-momentum:
natural observers are GR-{\it inertial},
on their worldlines 
$\vec{\cal{E}}_{\rm} = 0,$ $\vec{\cal{B}}_{\rm} = 0,$
and there is {\it no repulsive gravity}.

But for the {\it accelerated expansion}
of our {\it strongly perturbed} late universe 
measured on our past light-cone,
the natural observers, FIDOs, 
are everywhere on the light rays between the supernova explosion
at redshift $z < 2$ and us at $z = 0$
as defined in Sect.~\ref{subsect.FIDOs}.
Almost all of these FIDOs are GR-noninertial
and measure contributions from {\it repulsive gravity}.

\subsection{Gravito-Amp\`ere law of General Relativity
\label{gravito-Ampere}}

For a primary observer,
his local {\it frame} of LONBs
gives his Ricci LONB-connections
$(\omega_{\hat{a}\hat{b}})_{\hat{c}}$,
Eqs.~(\ref{Ricci.connections.spat.local.frame}),
in his time direction and his 3-directions.
The Ricci LONB-connections determine
$R^{\,\hat{i}}_{\, \, \, \, \hat{0}}$,
which is given by the three terms in the Golden Rule,
Eqs.~(\ref{Cartan.wedge}-\ref{Cartan.d.e}),
\begin{eqnarray}
  (R^{\, \hat{i}}_{\, \, \, \hat{0}})^{(\partial \tilde{\omega})}
  &=& -\,\partial_{\hat{k}}(\omega^{\hat{i}}_{\, \, \, \hat{k}})_{\hat{0}}
  \,=\,(\mbox{curl}\,
      \vec{\cal{B}}_{\rm g})^{\hat{i}},
      \nonumber
  \\
(R^{\, \hat{i}}_{\, \, \, \hat{0}})^{(\rm
\tilde{\omega}\wedge\tilde{\omega})}
  &=&  - (\omega^{\hat{i}}_{\, \, \, \hat{0}})_{\hat{k}}
      (\omega^{\hat{0}}_{\, \, \, \hat{k}})_{\hat{0}}
  =  -\,(\vec{\cal{E}}_{\rm g} \times
      \vec{\cal{B}}_{\rm g})^{\hat{i}},
      \nonumber
  \\
  (R^{\, \hat{i}}_{\, \, \, \hat{0}})^{(\partial\tilde{e})}
    &=& -  (\omega^{\hat{i}}_{\, \, \, \hat{k}})_{\hat{0}}
       (\omega^{\hat{0}}_{\, \, \, \hat{k}})_{\hat{0}} 
  = -\,(\vec{\cal{E}}_{\rm g} \times
      \vec{\cal{B}}_{\rm g})^{\hat{i}}.
        \nonumber
\end{eqnarray}
The sum of these three terms give $R^{\, \hat{i}}_{\, \, \, \hat{0}}$
expressed by the
Ricci LONB-connections of the primary observer,
\begin{eqnarray}
  R^{\,\hat{i}}_{\,\,\,\hat{0}}
  \,&=&\,
          (\mbox{curl}\,\vec{\cal{\cal{B}}}_{\rm g})^{\hat{i}}
         \, - \,2\,
        (\vec{\cal{E}}_{\rm g}\times\vec{\cal{\cal{B}}}_{\rm g})^{\hat{i}}.
        \label{Ri0.curlB}
\end{eqnarray}
The simplicity of Eq.~(\ref{Ri0.curlB}) depends 
on our frame of a 
noninertial observer
  with its LONB connections,  
  Eqs.~(\ref{Ricci.connections.spat.local.frame}).

The Ricci $R^{\,\hat{i}}_{\, \, \,  \hat{0}}$
equation of General Relativity states,
\begin{eqnarray}
  R^{\,\hat{i}}_{\,\,\,\hat{0}}
  \,&=&\,-\,8\pi G\, (J_{\varepsilon}^{\hat{i}})_{\rm matter+EM}.
        \nonumber
\end{eqnarray}
Eliminating $R^{\,\hat{i}}_{\,\,\,\hat{0}}$ in the last
two equations for~$R^{\,\hat{i}}_{\,\,\,\hat{0}}$ gives,
\boldmath
\begin{eqnarray}
  &&\mbox{\bf gravito-Amp\`ere law of General Relativity}
     \nonumber
  \\
  &&\mbox{\bf in local frame of GR-noninertial observer}
     \nonumber
     \\
  &&\mbox{\bf curl}\,\vec{\cal{B}}_{\rm g}
   = -8\pi G(\vec{J}_{\varepsilon})_{\rm matter}
      \nonumber
  \\
  &&\quad\quad\quad\quad\,\,\,\,
     -2G(\vec{E}\times\vec{B})_{\rm EM}
     +2(\vec{\cal{E}}_{\rm g}\times\vec{\cal{B}}_{\rm g}).
     \label{gravito.Ampere}
\end{eqnarray}
\unboldmath
The gravito-Amp\`ere law of GR
has {\it no term} $\partial_{\hat{t}} \vec{\cal{E}}_{\rm g}$
in contrast to   
$\partial_t \vec{E}$
in the Maxwell-Amp\`ere law.
If the gravito-Amp\`ere law had a
term $\partial_{\hat{t}} \vec{\cal{E}}_{\rm g},$
gravitational vector waves, which do not exist, would be predicted.

The  form of bilinear expressions
in~$(\vec{\cal{E}}_{\rm g}, \vec{\cal{B}}_{\rm g})$
follows from~$J^P$ covariance: 
sources of~$\mbox{curl}\,\vec{\cal{B}}_{\rm g}$ have~$J^P = 1^-$.

\subsubsection{Gravitational energy current density
  $\vec{J}_{\varepsilon}$ 
\protect\\
  depends on observer's local frame}

The {\it sources} of $\mbox{curl}\,\vec{\cal{B}}_{\rm g}$
in the gravito-Amp\`ere law are
the energy current density of (1)~matter, (2)~electromagnetic field
(Poynting vector),
\begin{eqnarray}
  \vec{J}_{\varepsilon}^{\,(\rm EM)} \,
  &=&\,(\vec{E}\times\vec{B})/(4\pi),
  \nonumber
\end{eqnarray}
plus (3)~a term which must be identified with
the energy current density of the gravitational field
(gravito-Poynting vector),  
\begin{eqnarray}
  (\vec{J}_{\varepsilon})_{\rm grav}\,
  &=&\, - (\vec{\cal{E}}_{\rm g}\times\vec{\cal{B}}_{\rm g})/(4\pi G).
  \label{grav.energy.current}
\end{eqnarray}
The gravito-Amp\`ere law can be written,
\boldmath
\begin{eqnarray}
  \mbox{\bf curl}\,\vec{\cal{B}}_{\rm g}\,
  &=&\,-\,8\pi G (\vec{J}_{\varepsilon})_{\rm matter+EM+grav.fields}.
\end{eqnarray}
\unboldmath

The gravitational energy current
density~$(\vec{J}_{\varepsilon})_{\rm grav}^{(P)}$
and~$(\mbox{curl}\,\vec{\cal{B}}_{\rm g})_P$
depend on the {\it observer} with worldline through $P$
with $\bar{u}_{\rm obs}^{(P)} = \bar{e}_{\hat{0}}^{(P)}$.~---
On the worldline of a GR-{\it inertial} observer 
and in his local frame $(\vec{J}_{\varepsilon})_{\rm grav}^{(P)} = 0. $

\subsection{Gravito-Faraday law of General Relativity
\label{grav.Faraday}}

The first {\it Bianchi identity}
for the Riemann
tensor $(R^{\,\hat{a}}_{\,\,\,\hat{b}})_{\hat{c}\hat{d}}$ states that   
the sum of the  cyclic permutations of the three lower
indices~$(\hat{b}, \hat{c}, \hat{d})$ gives zero,
\begin{eqnarray}
\sum_{{\rm cyclic}\, (\hat{b} \hat{c} \hat{d})} \, 
(R^{\, \hat{a}}_{\, \, \, \hat{b}})_{\hat{c} \hat{d}}  
\, \, &=& \, \, 0.
\nonumber
\end{eqnarray}
If two of the three cyclically permuted indices
$(\hat{b}, \hat{c}, \hat{d})$ are equal,
the  first Bianchi identity is empty.

The first Bianchi identity for {\it two} indices $\hat{0}$ is,   
\begin{eqnarray} 
\sum_{\rm cyclic\,lower\,ind.} \, 
(R^{\, \hat{0}}_{\,\,\,\hat{i}})_{\hat{0} \hat{j}}\,  
&=&\,\,(R^{\,\hat{0}}_{\,\,\,\hat{i}})_{\hat{0}\hat{j}}\,\,-\,\,
    [ \, \hat{i} \Leftrightarrow \hat{j} \, ] \, \, = \, \, 0.
    \nonumber
\end{eqnarray}

In the frame
of a primary GR-noninertial observer,
his Ricci LONB-connections
$(\omega_{\hat{a}\hat{b}})_{\hat{c}}^{(r = 0)}$ are given by 
Eq.~(\ref{Ricci.connections.spat.local.frame}).~---
The {\it Golden Rule}
(for the Riemann tensor in LONB-components
expressed by Ricci LONB-connections),
is given by three terms, Eqs.~(\ref{Cartan.wedge}-\ref{Cartan.d.e}).

The {\it derivative terms} $(\partial\,\tilde{\omega})$,  
Eq.~(\ref{Cartan.partial.omega}),
give,  
\begin{eqnarray}
&&(R^{\hat{0}}_{\,\,\,\hat{i}})_{\hat{0}\hat{j}}^{(\partial\tilde{\omega})} 
  \,=\,\partial_{\hat{0}} 
        (\omega^{\hat{0}}_{\,\,\,\hat{i}})_{\hat{j}}
       -\partial_{\hat{j}} 
        (\omega^{\hat{0}}_{\,\,\,\hat{i}})_{\hat{0}}
 \,=\,-\partial_{\hat{0}} {\cal{B}}_{\hat{i}\hat{j}} 
      +\partial_{\hat{j}} {\cal{E}}_{\hat{i}}, 
          \nonumber
  \\
  && (R^{\hat{0}}_{\,\,\,\hat{j}})_{\hat{i}\hat{0}}
     ^{(\partial\tilde{\omega})} 
  \,=\,\partial_{\hat{i}} 
       (\omega^{\hat{0}}_{\,\,\,\hat{j}})_{\hat{0}}
      -\partial_{\hat{0}}
      (\omega^{\hat{0}}_{\,\,\,\hat{j}})_{\hat{i}}
      \,=\,\,\,\,
      \partial_{\hat{0}} {\cal{B}}_{\hat{j}\hat{i}}
      - \partial_{\hat{i}} {\cal{E}}_{\hat{j}},
      \nonumber
\\
&&\sum_{\rm cyclic\,lower\,ind.}  
(R^{\,\hat{0}}_{\,\,\,\hat{i}})_{\hat{0}\hat{j}}^{(\partial\tilde{\omega})}  
\,=\,-\,(2\,\partial_{\hat{t}}\vec{\cal{B}}_{\rm g}
    \,+\,\mbox{curl}\,\vec{\cal{E}}_{\rm g}\,)_{\hat{i} \hat{j}}.
        \nonumber
\end{eqnarray}

The {\it wedge term} $(\tilde{\omega} \wedge \tilde{\omega})$ 
in the Golden Rule,
   Eq.~(\ref{Cartan.wedge}),
with the Ricci LONB-connections  of the primary observer,
$(\omega^{\hat{a}}_{\, \, \, \hat{b}})_{\hat{c}}^{(r = 0)}$,
Eq.~(\ref{Ricci.connections.spat.local.frame}),
gives a {\it vanishing} contribution to the Bianchi identity,
\begin{eqnarray}
\sum_{\rm cycl.lower\,ind.}  
  (R^{\,\hat{0}}_{\,\,\,\hat{i}})_{\hat{0}\hat{j}}^{(\tilde{\omega}
  \wedge\tilde{\omega})} 
    &=&
(\omega^{\hat{0}}_{\, \, \, \hat{k}})_{\hat{i}} 
         (\omega^{\hat{k}}_{\, \, \, \hat{j}})_{\hat{0}}
          - [ \, \hat{i} \Leftrightarrow \hat{j} \,] = 0.
      \nonumber
\end{eqnarray}

The $(d\tilde{e}^{\hat{s}})$ term
in the Golden Rule,
Eq.~(\ref{Cartan.d.e}),
also gives
a {\it vanishing} contribution to the Bianchi identity,
\begin{eqnarray}
  &&\sum_{\rm cycl.lower\,ind.} 
  (R^{\hat{0}}_{\,\,\,\hat{i}})_{\hat{0}\hat{j}}^{(d\tilde{e}^{\hat{s}})} 
  =
      \nonumber
      \\
  &&=\,(\omega^{\hat{0}}_{\,\,\,\hat{i}})_{\hat{s}} 
      [ (\omega^{\hat{s}}_{\,\,\,\hat{0}})_{\hat{j}}
      - (\omega^{\hat{s}}_{\,\,\,\hat{j}})_{\hat{0}}   ]
  - [ \hat{i}\Leftrightarrow\hat{j}]\,
   =\, 0.
       \nonumber
\end{eqnarray}

Conclusion: in the local frame of a primary observer
with his Ricci LONB-connections
$(\omega^{\hat{a}}_{\,\,\,\hat{b}})_{\hat{c}}^{(r=0)}$,
Eq.~(\ref{Ricci.connections.spat.local.frame}),
the sum of the three terms in the Golden Rule,
Eq.~(\ref{golden.rule.curvature}),
gives the  first Bianchi identity
with two indices $\hat{0}$:
\boldmath
\begin{eqnarray}
  &&\mbox{\bf gravito-Faraday law of General Relativity}
     \nonumber
  \\
  &&\quad \quad\quad\quad     
    \mbox{\bf curl} \, \vec{\cal{E}}_{\rm g}  \, + \, 
    2 \, \partial_{\hat{t}} \, \vec{\cal{B}}_{\rm g}\,=\,0.
\end{eqnarray}
\unboldmath
The  form of bilinear expressions
in~$(\vec{\cal{E}}_{\rm g}, \vec{\cal{B}}_{\rm g})$
follows from covariance under~$J^P$: 
$\mbox{curl}\,\vec{\cal{E}}_{\rm g}$ has $J^P = 1^+$,
therefore it cannot have a source bilinear
in~$(\vec{\cal{E}}_{\rm g}, \vec{\cal{B}}_{\rm g})$.

\boldmath
\subsection{Divergence of gravito-magnetic field
 $\vec{\cal{B}}_{\rm g}$ in GR 
\label{div.grav.magn}}
 \unboldmath

The first Bianchi identity for  one index $\hat{0}$
in the non-permuted position, 
$\,(R^{\,\hat{0}}_{\, \, \, \hat{i}})_{\hat{j} \hat{k}},\,$ 
states that the sum of the cyclic permutations 
of $(\hat{i} \hat{j} \hat{k})$ gives zero.

In the Golden Rule,
the {\it derivative term} $(\partial\,\tilde{\omega})$, 
Eq.~(\ref{Cartan.partial.omega}),
in our local {\it frame} of LONBs for a primary {\it observer},
   Eq.~(\ref{Ricci.connections.spat.local.frame}),
contributes,
\begin{eqnarray}
\sum_{{\rm cyclic} (\hat{i} \hat{j} \hat{k})}\, 
(R^{\, \hat{0}}_{\,\,\,\hat{i}})_{\hat{j}\hat{k}}^{(\partial\tilde{\omega})}  
  \,&=&\,
     \sum_{{\rm cyclic}(\hat{i} \hat{j} \hat{k})}\, 
 -2\,\partial_{\hat{i}} (\omega^{\hat{0}}_{\,\,\,\hat{j}})_{\hat{k}}\,=
\nonumber
\\
  =\,\sum_{{\rm cyclic}(\hat{i}\hat{j}\hat{k})}\,  
  2\,\partial_{\hat{i}} {\cal{B}}_{\hat{j}\hat{k}}^{(\rm g)}\, 
&=&\,\,2\,\mbox{div}\,\vec{\cal{B}}_{\rm g}. 
\nonumber
\end{eqnarray}

The {\it wedge term}
$(\tilde{\omega} \wedge \tilde{\omega})$
in the Golden Rule,
    Eq.~(\ref{Cartan.wedge}),
    contributes {\it zero}
in our {\it local frame},
    Eq.~(\ref{Ricci.connections.spat.local.frame}).

The term $(\partial\tilde{e}^{\hat{s}})$ in the Golden Rule,
   Eq.~(\ref{Cartan.d.e}),
in our local frame of the observer,
   Eq.~(\ref{Ricci.connections.spat.local.frame}), gives,
\begin{eqnarray}
\sum_{{\rm cyclic} (\hat{i} \hat{j} \hat{k})} \, 
  (R^{\, \hat{0}}_{\, \, \, \hat{i}})_{\hat{j} \hat{k}}
  ^{(d\tilde{e}^{\hat{s}})}  
  \,&=&\,\sum_{{\rm cyclic}(\hat{i} \hat{j} \hat{k})} \, 
          2 \, (\omega^{\hat{0}}_{\, \, \, \hat{i}})_{\hat{0}} \,
                  (\omega^{\hat{0}}_{\, \, \,  \hat{j}})_{\hat{k}}
\nonumber
\\
 =   \sum_{{\rm cyclic}(\hat{i} \hat{j} \hat{k})} \,  
2\,{\cal{E}}^{(\rm g)}_{\hat{i}} {\cal{B}}_{\hat{j} \hat{k}}^{(\rm g)} \, 
&=& \, 
   2\,\vec{\cal{E}}_{\rm g} \cdot \vec{\cal{B}}_{\rm g}. 
    \nonumber
\end{eqnarray}

The sum of the contributions gives,
\boldmath
\begin{eqnarray}
  &&\mbox{\bf div}\,\vec{\cal{B}}_{\rm g}\,\,
     \mbox{\bf law of General Relativity}
     \nonumber
    \\
      &&\quad \quad  
         \mbox{\bf div}\,\vec{\cal{B}}_{\rm g}\,+\, 
         \vec{\cal{E}}_{\rm g}\cdot\vec{\cal{B}}_{\rm g}\,=\,0.
\label{div.B.GR}
\end{eqnarray}
\unboldmath
The  form of bilinears 
in~$(\vec{\cal{E}}_{\rm g}, \vec{\cal{B}}_{\rm g})$
follows from covariance under~$J^P$: the source
of~$\mbox{div}\,\vec{\cal{B}}_{\rm g}$ has~$J^P = 0^-$.

The Bianchi identity with one index $\hat{0}$ 
in the cyclic permutation of the three lower indices is empty.

The gravito-Maxwell equations of GR
  for a chosen observer do not involve
        Einstein's $R_{\hat{i}\hat{j}}$ equations for  
        the intrinsic curvature of his 3-space.

\subsection{GR-effects from fictitious tidal accelerations
  \protect\\
  and Einstein's concepts of 1911
\label{R00.not.rel.accel}}

The acceleration of a freefalling particle is undefined
unless one has a {\it reference frame}:
Newton-inertial frames for Newton-gravity,
non-inertial frames in classical mechanics,
SR-inertial frames for Special Relativity,
the local frame of LONBs of a chosen {\it observer} for GR.

In this Subsection we show directly our crucial result:
the {\it acceleration-differences}
of neighboring freefalling particles spherically averaged
measured in the local frame of LONBs
of a GR-{\it noninertial observer}
is {\it not} given by the Ricci tensor,
but this acceleration-difference involves
$(\vec{\cal{E}}_{\rm g}^{\,2} + 2\vec{\cal{B}}_{\rm g}^{\,2}),$
which causes {\it repulsive gravity}.~---
Here we shall use only Einstein's concepts of 1911
without using Einstein's equations of 1915.

\subsubsection{Fictitious centrifugal acceleration
  in rotating frame\\
  gives repulsive Einstein-gravity
  from $\vec{\cal{B}}_{\rm g}^{\,2}$ 
\label{div.E.centrifugal}}

We give an elementary argument which shows that 
the acceleration-difference of freefalling particles
(measured in the frame of LONBs of a GR-{\it rotating observer}
and spherically averaged)
is {\it not} described by the Ricci $R^{\,\hat{0}}_{\,\,\hat{0}}.$

If at a point $P$ the sources of matter and electromagnetic fields
are zero, $(R^{\,\hat{0}}_{\,\,\,\hat{0}})^{(P)} = 0.$~---
But in a {\it rotating frame}
the centrifugal acceleration of classical mechanics
gives relative accelerations of quasistatic freefalling particles
which can be arbitrarily large,
\begin{eqnarray}
  \vec{a}_{\rm centrifugal}\,
  &=&\,\vec{\Omega}\times(\vec{r}\times\vec{\Omega}),
      \nonumber
  \\
  \mbox{div}\,\,\vec{a}_{\rm centrifugal}\,
  &=&\,2\,\vec{\Omega}^{\,2}.
      \nonumber
\end{eqnarray}
$\vec{\Omega}$ is the angular velocity of the frame-rotation
relative to Newton-inertial.~---
The classical Gauss law
for  accelerations of freefalling quasistatic particles
in a rotating frame is,
\begin{eqnarray}
  \mbox{div}\,\vec{a}_{\,\rm freefall}\,
  &=&\,-\,4\pi G\,\rho_{\rm mass}\,
      +\, 2\,\vec{\Omega}^{\,2}.
      \nonumber
\end{eqnarray}
Without Newton-inertial frames, one cannot separate
fictitious accelerations 
from classical gravitational accelerations.
Einstein's gravitational
acceleration~$\vec{\mathfrak{g}}_{\,\rm GR}$
is the sum of~$\vec{g}_{\,\rm Newton}$,
which is {\it generated} by {\it mass sources},
       and the {\it classical fictious acceleration}
       of the Newton-inertial frame relative
       to the chosen GR-observer, which is
       {\it not generated} by {\it mass sources},
       Eq.~(\ref{g.Einstein.minus.g.Newton}).~---
Einstein stated the {\it equivalence}
of classical {\it fictitious} accelerations 
with {\it additional gravitational} accelerations
in his approach.

In GR, the acceleration of quasistatic freefalling
particles relative to
the LONB of the chosen
{\it observer} (FIDO)
defines $\vec{\cal{E}}_{\rm g}^{\,(\rm GR)},$
Eq.~(\ref{def.grav.E.xxx}). 
The precession angular velocity
for spin-axes of comoving gyros 
relative to the LONB of the FIDO
defines $\vec{\cal{B}}_{\rm g}^{\,(\rm GR)}$,
Eq.~(\ref{def.B.g.x}).
This completes the transcription of the Gauss law
of Newton gravity
to acceleration-differences of freefalling particles
in the {\it frame} of a {\it rotating-freefalling} FIDO
with Einstein's concepts of 1911,
\begin{eqnarray}
  \mbox{div}\,\vec{\cal{E}}_{\rm g}\,
  &=&\,-\,4\pi G\,\rho_{\rm mass}\,
       +\,2\,\vec{\cal{B}}_{\rm g}^{\,2}.
\end{eqnarray}
 Einstein: 
``We are able to produce a  gravitational field
  merely by changing the system of reference''.

  We have given an elementary re-derivation
  of {\it repulsive gravity}
from the term~$2 \vec{\cal{B}}_{\rm g}^{\,2}$
in the gravito-Gauss law of GR 
for the 
frame of a freefalling-{\it rotating observer}.
We have used Einstein's concepts of 1911
without using any concepts of curved space-time.

The {\it acceleration-difference}
of freefalling particles depends on the observer
(GR-inertial versus GR-noninertial) with his reference frame.~---
In contrast, the {\it Riemann tensor} does not depend on
whether the observer is GR-inertial or not.

GR-texts, e.g.~\cite{Weinberg, Wald, Poisson.Will, Hartle}, imply:
if the Riemann tensor at~$P$ is zero,
the {\it acceleration-difference} of freefalling particles 
measured in the frame of a 
freefalling-{\it rotating observer} at~$P$ is also zero.
This statement is wrong, as shown by
the counter-example given here.

\newpage

\subsubsection{Fictitious repulsive tidal acceleration
  \protect\\
  in accelerated frame of Special Relativity
\label{repulsion.acc.frame.SR}}

Relative to a linearly
{\it accelerated observer} (nonrotating)
with his local reference frame of LONBs,
we derive our new fictitious repulsive
{\it acceleration-difference}
of {\it freefalling} particles in Special Relativity.~---
In the gravito-Gauss law of General Relativity,
this acceleration difference gives
the repulsive~$\vec{\cal{E}}_{\rm g}^{\,2}$ term,
which contributes to
the accelerated expansion of the universe.

In the local frame of an observer at $x = 0$,
non-rotating but accelerated relative to inertial
in the positive $x$-direction,
two freefalling particles are initially at rest 
at~$x^i_A = 0$ and~$x^i_B = (\delta x, 0, 0)$.

For first-order~$\delta t$, the freefalling particles 
have~$\delta s \propto (\delta t)^2 = 0$ in the observer's frame.
Their velocities are first order infinitesimal,
hence no length contraction and no time dilation.
The velocity-difference of the particles remains zero
in the SR-noninertial frame.~---
But we show that their {\it acceleration-difference}  is {\it nonzero}
in the reference frame of the accelerated observer.

Relative to the SR-inertial time-axis,
the observer's time-axis 
gets tilted more and more (as time goes on)
in the positive $x$-direction
relative to the SR-inertial time-axis.
The observer's $x$-axis gets tilted
relative to the SR-inertial $x$-axis
in the positive time direction.

The {\it time-lapse} $\alpha$ in the local accelerated frame
is the measured time elapsed
between two equal-time
slices~$\Sigma_t$ 
of the primary accelerated observer
separated by unit time on his wristwatch.~---
The lapse~$\alpha$ at the position of particle~$A$ is
$\alpha_{A} = 1.$ The lapse at $B$ is,
\begin{eqnarray}
  \alpha_B \,
  &=&\,1 - ({\mathfrak{a}}^{\hat{x}}_{\,A})_{\rm rel.to\,obs}
     \,\, \delta \hat{x}_B.
      \nonumber
\end{eqnarray}
In our example, the
acceleration~$({\mathfrak{a}}^{\hat{x}}_{\,A})_{\rm rel.to\,obs}$
is negative, hence $\alpha_B > 1.$~---
The acceleration of freefall particle $B$
relative to the observer is,
\begin{eqnarray}
  {\mathfrak{a}}^{\hat{x}}_{\,B}\,
  &=&\, {\mathfrak{a}}^{\hat{x}}_{\,A}\,\,\alpha_B^{-1}\,\,
      =\,\,{\mathfrak{a}}^{\hat{x}}_{\,A}\,
      (1 + {\mathfrak{a}}^{\hat{x}}_{\,A}
     \,\, \delta \hat{x}_B).
      \nonumber
\end{eqnarray}

The {\it difference of accelerations}    
of our neighbouring freefalling particles 
in the frame of the {\it accelerated} and nonrotating {\it observer}
in Special Relativity is nonzero,
\begin{eqnarray}
\partial_{\hat{x}}\, {\mathfrak{a}}^{\hat{x}}_{\,B}\,
  &=&\,({\mathfrak{a}}^{\hat{x}}_{\,A})^2.
\label{accel.diff.nonzero.SR}
\end{eqnarray}
This shows: 
the {\it acceleration-difference}
of neighbouring {\it free-falling} particles,
$\partial_{\hat{x}} {\mathfrak{a}}^{\hat{x}},$
measured in the local   
frame of
an {\it accelerated}-nonrotating {\it observer}
is nonzero in Special Relativity, it is
{\it not given} by the {\it Riemann tensor}:
the contrary conclusions 
in~\cite{Wald,
  Poisson.Will}
are wrong.

For Special Relativity in an accelerated-nonrotating frame,
the {\it acceleration-difference}
of infinitesimally separated {\it freefalling particles},
$\partial_{\hat{x}}\, {\mathfrak{a}}^{\hat{x}}_{\rm ff}$,
is {\it positive}, ``repulsive''.

The gravito-electric field~$\vec{\cal{E}}_{\rm g}$
measured in the local frame of an accelerated-nonrotating observer
is defined equal to the acceleration~$\vec{\mathfrak{a}}$
{\it relative} to the {\it observer}
of freefalling particles (released by the observer).
The last equation gives
the acceleration-difference of neighboring freefalling particles
spherically averaged. 
With  $ c \neq 1,$
\begin{eqnarray}
  \mbox{div}\,\vec{\cal{E}}_{\rm g}\,
  &=&\,\frac{1}{c^2}\,\vec{\cal{E}}_{\rm g}^{\,2}.
      \label{grav.Gauss.SRT}
\end{eqnarray}
This is a law of Special Relativity:
for accelerations of Newtonian order of magnitude,
it reduces to~$\mbox{div}\,\vec{\cal{E}}_{\rm g}\ = 0.$

Eq.~(\ref{grav.Gauss.SRT}) is our gravito-Gauss law of GR
in the {\it reference frame}
of an {\it accelerated-nonrotating observer}
at a point free of matter and electromagnetic fields.
We have re-derived this law using Einstein's concepts of 1911,
but without using Einstein's equations.


\subsubsection
{Derivation of gravito-Faraday law of GR
  \protect\\
  from Euler’s fictitious acceleration
  \protect\\
  with Einstein's concepts of 1911}

We have derived the gravito-Faraday law of GR
using the tools of GR in
   Sect.~\ref{grav.Faraday}.~---
We now re-derive this gravito-Faraday law 
with Einstein's concepts of 1911,
but without using Einstein's equations of 1915.

In a frame with time-dependent angular velocity $\vec{\Omega}$
relative to Newton-inertial,
the fictitious Euler acceleration of freefalling particles is,
\begin{eqnarray}
  (\vec{a}_{\rm Euler})_{\rm ff}^{(\rm rel.Euler-frame)}
  \,&=&\,\vec{r} \times \frac{d}{dt}\,
        \vec{\Omega}_{\,\rm Euler\,frame}^{\,(\rm rel.Newton-inert)}.
  \nonumber
  \end{eqnarray}

  Euler's frame is rotating relative to
  the Newton-nonrotating frame.
  But the Euler acceleration also holds for the frame rotating
  relative to local Einstein-nonrotating frames.

  The operational definitions
  of~$(\vec{\cal{E}}_{\rm g}, \vec{\cal{B}}_{\rm g})$
  by the gravitational acceleration of freefalling particles
  and the gravitational precession of gyros
  relative to the observer are, 
   \begin{eqnarray} \vec{a}_{\rm ff}^{\,(\rm relat.to\,obs.LONB)}\, 
    &\equiv&\,\vec{\cal{E}}_{\rm g},
                \nonumber
     \\
   \vec{\Omega}_{\rm gyro}^{\,(\rm relat.to\,obs.LONB)}
     \,&\equiv&\,-\,\vec{\cal{B}}_{\rm g}.
                \nonumber
  \end{eqnarray}              
  \\
  $\vec{\Omega}_{\rm Euler\,frame}^{(\rm rel.to\,Newton-inert)}$
  and $\vec{\Omega}_{\rm gyro}^{\,(\rm relat.to\,obs.LONB)}$
  have opposite signs.
  Inserting the definitions of
  $(\vec{\cal{E}}_{\rm g}, \vec{\cal{B}}_{\rm g})$
  into Euler's fictitious acceleration 
  for the observer's frame gives,    
  \begin{eqnarray}\vec{\cal{E}}_{\rm g}\,
    =\,\vec{r}\times\partial_{\hat{t}}\vec{\cal{B}}_{\rm g},\,
&\quad&\, 
    \mbox{curl}\,\vec{\cal{E}}_{\rm g}\,
    =\,-\,2\,\partial_{\hat{t}}\vec{\cal{B}}_{\rm g}.
        \nonumber
  \end{eqnarray}
  This is the re-derivation of the gravito-Faraday law of GR
from Euler’s fictitious acceleration
using Einstein's concepts of 1911,
but without using Einstein's equations.


\begin{acknowledgments}
We thank  
J.~Fr\"ohlich,
G.M.~Graf,
J.~Hartle,
N.~Straumann,
K.~Thorne,
and R.~Wald
for helpful discussions.
\end{acknowledgments}





\end{document}